\documentclass[useAMS,usenatbib]{mn2e}

\usepackage{amssymb}
\usepackage{amsmath}
\usepackage[dvips]{graphicx}
\usepackage{epsfig}
\usepackage{multicol}

\title[GRB plateau correlations favour thick shells over thin]{Gamma-ray burst afterglow plateau break time - luminosity correlations favour thick shell models over thin shell models}

\author[H.J. van Eerten]{Hendrik van Eerten$^1$\thanks{Alexander von Humboldt Fellow} \thanks{e-mail: hveerten@mpe.mpg.de} \\
$^{1}$Max-Planck-Institut f\"ur Extraterrestrische Physik, Giessenbachstra\ss e 1, 85748 Garching, Germany}

\begin{document}


\pagerange{\pageref{firstpage}--\pageref{lastpage}}

\maketitle

\label{firstpage}

\begin{abstract}
A number of correlations between observables have been found to exist for gamma-ray burst (GRB) afterglows, linking ejecta energy to prompt and afterglow energy release and linking early stage optical and X-ray luminosity to the end times of these stages. Here, these correlations are compared to thick and thin shell models for GRB afterglows. In the thick shell model, the time evolution of the underlying relativistic blast wave is still influenced by the original ejecta, while in the thin shell model most energy in the explosion has been transferred to the external medium. It is shown here that the observed correlations rule out basic thin shell models but not the basic thick shell model.    In the thick shell case, both forward shock and reverse shock dominated outflows are shown to be consistent with the correlations, using randomly generated samples of thick shell model afterglows.
\end{abstract}


\section{Introduction}
\label{introduction_section}

In no small part due to the launch of the \emph{Swift} satellite about ten years ago \citep{Gehrels2004}, the amount of high quality, early time gamma-ray burst (GRB) afterglow data has increased considerably. The \emph{Swift} era has revealed new features that pose additional constraints on theoretical models, such as X-ray plateaus lasting up $10^{3-4}$ seconds for long GRBs \citep{Nousek2006, ZhangBing2006}, where the emission decays more slowly than expected for a decelerating afterglow blast wave. A plateau provides at least a flux level, light curve slope and a turnover time to normal light curve decay that need to be accommodated by any valid model.
In addition, more early time optical afterglow data are becoming available from Swift-UVOT and ground based observatories, revealing the existence of a separate early stage in the light curves in these bands as well (see e.g. \citealt{PanaitescuVestrand2008, PanaitescuVestrand2011, Filgas2011, Li2012} for examples), and again implying additional afterglow blast wave evolution in addition to late time deceleration.
Optical and X-ray early stages might not necessarily lie in the same spectral regime and therefore yield different constraints. Additionally, a number of recent studies report a series of correlations between various early stage parameters and other burst parameters that might serve to confirm or invalidate our previous notions about the GRB and afterglow mechanism \citep{Dainotti2008, Dainotti2010, Dainotti2011, PanaitescuVestrand2011, Li2012, Dainotti2013, Grupe2013, Margutti2013}.

GRB afterglows are expected to be produced by non-thermal emission from highly relativistic outflows. For massive relativistic ejecta, two categories of models can traditionally be identified: those with a \emph{thin} shell and those with a \emph{thick} shell \citep{SariPiran1995, Kobayashi1999, KobayashiSari2000}. An afterglow blast wave shell is considered thin if its initial width is so small that it quickly ceases to leave an imprint on the ejecta dynamics, which will then be dictated by the current ejecta radius and Lorentz factor instead. Specifically, this will occur before the reverse shock (RS), generated by the impact between ejecta and environment and running back into the ejecta, becomes relativistic. For thick shells, the RS will become relativistic during crossing of the ejecta and this will alter the ejecta dynamics.

In the collapsar scenario \citep{Woosley1993, MacFadyen1999}, the GRB is the result of the collapse of a massive star into a black hole. In this case, there is no clear mechanism to power outflows for $10^4$ seconds. Unless the ejecta are emitted with a range of Lorentz factors, where slower shells will fall behind faster shells initially before catching up (see e.g. \citealt{Nousek2006, GranotKumar2006}), the initial width of the ejecta will therefore be set by the size of the progenitor system or speed of light $c$ times the duration of the prompt emission. Combined with the ultra-high Lorentz factors that are typically inferred from the prompt emission, this naturally leads to a thin shell scenario where the shell starts to decelerate around $10^2$ s. (in the observer frame), and no plateau-type deviation from a standard decelerating shell afterglow light curve is expected. 

However, this also assumes that the initial Lorentz factor of the ejecta responsible for the afterglow emission is that of the outflow generating the prompt emission, which is not necessarily the case. For example, the production of a massive slower moving shell (``cocoon'') around the prompt emission outflow is a natural by-product of collapsar jet breakout. This cocoon is expected to be only mildly relativistic (see e.g. \citealt{RamirezRuiz2002, ZhangWeiqun2003, Morsony2007}). While still a thin shell in the previously defined sense, this would lead to an observer frame deceleration time around $10^4$ seconds, similar to the end time of the plateau. Two-component jet models (e.g. \citealt{RamirezRuiz2002, Peng2005, Granot2006}) therefore provide a natural candidate to explain afterglow plateaus (for optical and X-rays examples, see e.g. \citealt{Berger2003, Filgas2011}). Alternatively, the Lorentz factor of the blast wave could have dropped considerably early on due to a high mass density immediately surrounding the progenitor but not extending sufficiently far outward to impact the integrated column density at radii where the majority of afterglow emission takes place (and therefore not affecting the inferred values for afterglow densities from broadband modeling).

Other favoured explanations for afterglow plateaus include some form of energy injection into the jet. These can lead to thick shell-type scenario's where the width of the shell is set by the duration of the energy injection, be it through a continuum of sufficiently energetic shells with decreasing Lorentz factor entering the reverse shock or through a continuing source luminosity. A leading candidate for the source of the injection of energy of the latter type is a magnetar, an extremely magnetic and (temporarily) stable neutron star formed at the moment of collapse, that sheds its rotational energy \citep{DuncanThompson1992, Usov1992, Dai1998, ZhangMeszaros2001}.

Finally, there are explanations for differing early time afterglow behavior that do not include altering the jet dynamics. Examples of these include time evolution of the microphysics parameters \citep{Granot2006} and viewing angle effects \citep{EichlerGranot2006}.

In this study, I discuss the implications of the separate correlations in optical and X-rays between early stage end time $T$ and X-ray and optical luminosities $L_X$ and $L_O$ at this time, and the absence of a clear correlation between break time and total energy, for thick and thin shell scenario's of GRB afterglows. The relevant (non-)correlations are described in section \ref{key_correlations_section}. In section \ref{model_implications_section}, the implications of the correlations for the thick and thin shell models are described, and reverse shock emission in a thick shell scenario is found to be favoured in theory. Section \ref{synthetic_section} explores these implications for randomly sampled synthetic light curves generated from reasonable underlying distributions of the model parameters. In practice, both reverse and forward shock thick shell emission are found to be consistent with the correlations, with the preference for reverse shock emission not sufficient to overcome the noise level in the statistics. Thin shells models remain ruled out. Section \ref{discussion_section} closes off with a summary and additional discussion.

\section{Some key (non-)correlations}
\label{key_correlations_section}

GRB afterglow light curves and prompt emission can be described by two types of parameters, those that are either directly observable (e.g. plateau phase end time $T$) or those inferred based on some underlying model (e.g. inferred isotropic equivalent energy of the ejecta $E_{iso}$). Over the past years various groups have reported on the existence and absence of correlations between various parameters. One correlation that was reported early on \citep{Frail2001, PanaitescuKumar2001}, is that between isotropic equivalent energy release in gamma rays $E_{\gamma,iso}$ and $E_{iso} (t \ge T)$,
\begin{equation}
E_{\gamma, iso} \propto E_{iso}(t \ge T).
\label{E_gamma_E_iso_correlation_equation}
\end{equation}
Once \emph{Swift} revealed the existence of plateaus, raising the possibility of prolonged injection of energy, $E_{iso}$ could no longer be assumed constant throughout the entire light curve evolution. However, correlation \ref{E_gamma_E_iso_correlation_equation} holds also at a specific early time $t_c$ (the ``deceleration'' time, see e.g. \citealt{SariPiran1995, PanaitescuKumar2000} and the discussions in \citealt{Granot2006, ZhangBing2007}).

A correlation between two observable quantities, X-ray luminosity $L_X (T)$ (erg s$^{-1}$) and plateau break time $T$, has been reported by \cite{Dainotti2008, Dainotti2010, Dainotti2011, Dainotti2013, Margutti2013}:
\begin{equation}
L_{X} (T) \propto T^{-(1.07^{+0.20}_{-0.09})}.
\label{LTX_correlation_equation}
\end{equation}
hereafter referred to as the ``LTX correlation''. Both \cite{Dainotti2013} and \cite{Margutti2013} study large samples, and the former perform a detailed analysis of potential intrinsic redshift-based biases and redshift-induced observational biases as well. The results from both studies are consistent within their 1$\sigma$ errors. In the equation above we list the best fit value from \cite{Dainotti2013}, but increase the error bars to include the outer range from \cite{Margutti2013}, to be on the safe side. The correlation is stronger in the rest frame of the burster, and eq. \ref{LTX_correlation_equation} is expressed in this frame. In the remainder of this manuscript, all quantities are expressed in this rest frame. Fixed redshift $z = 0$ and fixed luminosity distance $d_L$ are used when quantities are generated from thick and thin shell analytical models, allowing us to ignore redshift effects and to conflate luminosity $L$ and (monochromatic) flux $F$ (erg s$^{-1}$ cm$^{-2}$; erg s$^{-1}$ cm$^{-2}$ Hz$^{-1}$ if monochromatic) as far as correlations are concerned. The LTX correlation and its optical counterpart, discussed below, are assumed to be free from observational and redshift biases (see \citealt{Dainotti2010, Dainotti2011, Dainotti2011systematics, Dainotti2013}). The consistency between \cite{Margutti2013} and \cite{Dainotti2013}, and the discussion by the latter authors, support the notion that these biases would not be problematic even when not explicitly accounted for. We will, however, further ignore the correlation between X-ray flux and observer plateau end time reported by \cite{Margutti2013}, where the imprint of redshift skews the slope.

A similar correlation between optical (R-band) luminosity $L_R (T)$ and $T$ (``LTO correlation'') at early times is reported by \cite{PanaitescuVestrand2011, Li2012}:
\begin{equation}
L_O (T) \propto T^{-0.78 \pm 0.08},
\label{LTO_correlation_equation}
\end{equation}
using the error bars from \cite{Li2012} (the sample from \citealt{PanaitescuVestrand2011} is significantly smaller. In the current study, I will show their error range separately in plots). The end times for this early optical phase and for the early X-ray plateau phase were found be roughly consistent \citep{Li2012}. As is the case for X-ray light curves, where plateaus are seen in roughly one third of the Swift XRT sample \citep{Liang2007, Evans2009, Racusin2009, Margutti2013}, not all optical light curves include an early shallow decay stage (\citealt{Li2012} report 39 out of 146). In both X-rays and optical, the presence of flaring behavior can render precise determination of underlying early time features more difficult. Interestingly, the slope of the optical correlation is different from that in X-rays even accounting for the error bars in eqs. \ref{LTX_correlation_equation} and \ref{LTO_correlation_equation}. This strongly suggests that we are \emph{not} looking at the same spectral regime in X-rays and optical, consistent with earlier studies comparing optical and X-ray emission for individual bursts (see e.g. \citealt{Greiner2011}).

\cite{Li2012} further report that they did not find a correlation between $T$ and $E_{R, iso}$, the total $R$-band energy release in the plateau phase from 10 s to $T$, while a rough proportionality was observed between $E_{R, iso}$ and $E_{\gamma, iso}$. Similarly, \cite{Margutti2013} report correlations between the energy release in X-rays $E_{X, iso}$ and $E_{\gamma, iso}$, regardless of whether $E_{X, iso}$ is calculated for the entire light curve or just the plateau phase, as well confirming the absence of a correlation between $T$ and $E_{X, iso}$ or $E_{\gamma, iso}$ reported by \cite{Dainotti2011}. In view of eq. \ref{E_gamma_E_iso_correlation_equation}, the results from \cite{Li2012}, \cite{Dainotti2011} and \cite{Margutti2013} imply that $E_{iso}$ and $T$ are uncorrelated.

\section{Model implications of correlations}
\label{model_implications_section}

\subsection{Basic Thin shells}

In a thin shell two-jet or jet-cocoon interpretation of the plateau phase, the observed break time $T$ follows from the deceleration radius of the broad, slow jet and is given by \citep{SariPiran1995, Yi2013}:
\begin{equation}
T \propto \left( \frac{E_{iso}}{n_{ref} R_{ref}^k m_p c^2} \right)^{1/(3-k)} / (\eta^{2(4-k)/(3-k)} c).
\label{T_equation}
\end{equation}
Here $m_p$ is the proton mass, $n_{ref}$ the number density of the circumburst medium at reference distance $R_{ref}$, $k$ the circumburst density radial slope (i.e. $n \equiv n_{ref} (r / R_{ref})^{-k}$) and $\eta$ the initial Lorentz factor of the ejecta. The thin shell model therefore predicts a clear correlation between $E_{iso}$ and $T$, e.g. $T \propto E_{iso}^{1/3}$ for a homogeneous medium with $k = 0$, rather than the reported non-correlations. This provides a strong argument against thin shell models, although at this point it is still possible that, in practice, a correlation of this type remains buried in the noise for a sample of afterglow light curves drawn from broad underlying distributions of variables such as $E_{iso}$, $n_{ref}$. We will return to this in the next section.

In a standard synchrotron emission model \citep{Wijers1997, Sari1998, Granot2002}, the spectrum is characterized (across the frequencies under consideration) by a peak flux $F_{peak}$, a characteristic break frequency $\nu_m$ associated with the lower Lorentz factor boundary on the shock-accelerated electron population and a characteristic frequency $\nu_c$ associated with the electron Lorentz factor beyond which the cooling time becomes short enough to become noticeable across the ejecta. Following \cite{Granot2002, vanEerten2009} for the labeling of the indices $D$, $E$, $F$, $G$, $H$, the various power law components of the spectrum potentially observable at frequency $\nu$, can be summarized as:
\begin{eqnarray}
F_D \equiv F_{peak} (\nu / \nu_m)^{1/3} & : & \nu < \nu_m < \nu_c, \nonumber \\
F_E \equiv F_{peak} (\nu / \nu_c)^{1/3} & : & \nu < \nu_c < \nu_m, \nonumber \\
F_F \equiv F_{peak} (\nu / \nu_c)^{-1/2} & : & \nu_c < \nu < \nu_m, \nonumber \\
F_G \equiv F_{peak} (\nu / \nu_m)^{(1-p)/2} & : & \nu_m < \nu < \nu_c, \nonumber \\
F_H \equiv F_{peak} (\nu_c / \nu_m)^{(1-p)/2} (\nu / \nu_c)^{-p/2} & : & \nu_m, \textrm{ } \nu_c < \nu,
\label{spectral_regime_equations}
\end{eqnarray}
where $p$ is the power law slope of the shock-accelerated electron population, such that electron number density $n_e$ depends on electron Lorentz factor $\gamma_e$ according to $n_e(\gamma_e) \propto \gamma_e^{-p}$.

For thin shells, the deceleration time, the moment when the RS crosses the ejecta and the point when the RS becomes relativistic, all occur approximately at $T$ and together mark the end of the plateau. With $T$ itself a function of $E_{iso}$, $\eta$ and $n_{ref}$, the luminosities at this point in time can be expressed as \citep{Yi2013, Gao2013}:
\begin{eqnarray}
L_D & \propto & (\eta^{\frac{3+k}{3(3-k)}} \textrm{ or } \eta^{\frac{2(k-2)}{3-k}}) \, n_{ref}^{\frac{1}{3-k}} E_{iso}^{\frac{9-4k}{3(3-k)}} , \nonumber \\
L_E & \propto & (\eta^{\frac{5+3k}{3(3-k)}} \textrm{ or } \eta^{\frac{2(3k-2)}{3(3-k)}}) \,  n_{ref}^{\frac{7}{3(3-k)}} E_{iso}^{\frac{11-6k}{3(3-k)}}  , \nonumber \\
L_F & \propto & (\eta^{\frac{3}{3-k}} \textrm{ or } \eta^{\frac{k}{(3-k)}}) \,  n_{ref}^{\frac{3}{2(3-k)}} E_{iso}^{\frac{3(2-k)}{2(3-k)}}  , \nonumber \\
L_G & \propto & (\eta^{\frac{12-k+6p-pk}{2(3-k)}} \textrm{ or } \eta^{\frac{12-k+12p-3pk}{2(3-k)}}) \,  n_{ref}^{\frac{3(3+p)}{4(3-k)}} E_{iso}^{\frac{12-7k-pk}{4(3-k)}}  , \nonumber \\
L_H & \propto & (\eta^{\frac{4-2k+6p-pk}{2(3-k)}} \textrm{ or } \eta^{\frac{2k-8+12p-3pk}{2(3-k)}}) \,  n_{ref}^{\frac{3p-2}{4(3-k)}} E_{iso}^{\frac{8-2k-pk}{4(3-k)}}. \nonumber
\end{eqnarray}
Emission from the RS and FS regions differ only in their $\eta$-dependencies. In the above, the first option refers to RS, the second to FS emission. 
There is no spectral regime for which the $E_{iso}$, $n_{ref}$ and $\eta$ dependencies of $T$ can be used to reduce the flux to a function of $T$ only. It follows that no LTX / LTO type correlations (or any correlations at all) emerge from a basic thin shell model where the physics parameters are free to vary.

\subsection{Extensions to the basic thin shell model}

\begin{table}
\centering
\begin{tabular}{rrrrr}
 & & $w$ & $r$ & $s$ \\
\hline
$k = 0$  & D & $0.02^{+0.05}_{-0.02}$ & $1.86^{+0.29}_{-0.14}$ & $[-33.93, 342.00]$ \\ 
RS  & E & $-0.27^{+0.05}_{-0.02}$ & $1.46^{+0.27}_{-0.13}$ & $5.46^{+2.53}_{-0.89}$ \\ 
 & F & $-0.11^{+0.05}_{-0.03}$ & $1.37^{+0.31}_{-0.15}$ & $12.93^{+18.20}_{-3.62}$ \\ 
 & G & $-0.70^{+0.08}_{-0.04}$ & $-0.99^{+0.42}_{-0.20}$ & $-1.43^{+0.50}_{-0.20}$ \\ 
 & H & $-0.03^{+0.06}_{-0.03}$ & $-0.01^{+0.49}_{-0.24}$ & $[-13.00, -4.47 ]$ \\ 
\hline
$k = 0$  & D & $0.02^{+0.05}_{-0.02}$ & $3.09^{+0.23}_{-0.11}$ & $[ -52.44, 592.00]$ \\
FS  & E & $-0.27^{+0.05}_{-0.02}$ & $2.09^{+0.24}_{-0.11}$ & $7.83^{+2.98}_{-1.05}$ \\
 & F & $-0.11^{+0.05}_{-0.03}$ & $1.73^{+0.29}_{-0.14}$ & $16.42^{+21.23}_{-4.23}$ \\
 & G & $-0.70^{+0.08}_{-0.04}$ & $-2.61^{+0.50}_{-0.24}$ & $-3.76^{+0.32}_{-0.13}$ \\
 & H & $-0.03^{+0.06}_{-0.03}$ & $-0.21^{+0.50}_{-0.25}$ & $-8$ \\
\hline
$k = 2$  & D & $0.05^{+0.12}_{-0.07}$ & $1.86^{+0.27}_{-0.15}$ & $[-12.64, 112.67]$ \\ 
RS  & E & $-1.71^{+0.58}_{-0.38}$ & $0.83^{+0.68}_{-0.44}$ & $0.49^{+0.84}_{-0.30}$ \\ 
 & F & $-0.40^{+0.22}_{-0.13}$ & $1.20^{+0.44}_{-0.26}$ & $2.98^{+6.07}_{-1.21}$ \\ 
 & G & $5.34^{+2.63}_{-0.63}$ & $9.66^{+3.43}_{-0.82}$ & $-1.81^{+0.17}_{-0.07}$ \\ 
 & H & $-0.08^{+0.19}_{-0.11}$ & $-0.12^{+0.70}_{-0.42}$ & $[-5.67, -2.82]$ \\ 
\hline
$k = 2$  & D & $0.05^{+0.12}_{-0.07}$ & $3.05^{+0.12}_{-0.07}$ & $[-18.81, 196.00 ]$ \\
FS  & E & $-1.71^{+0.58}_{-0.38}$ & $2.19^{+0.39}_{-0.25}$ & $1.28^{+0.99}_{-0.35}$ \\
 & F & $-0.40^{+0.22}_{-0.13}$ & $2.60^{+0.22}_{-0.13}$ & $6.47^{+9.10}_{-1.81}$ \\
 & G & $5.34^{+2.63}_{-0.63}$ & $13.81^{+5.95}_{-1.42}$ & $-2.59^{+0.11}_{-0.04}$ \\
 & H & $-0.08^{+0.19}_{-0.11}$ & $-0.33^{+0.74}_{-0.44}$ & $-4$ \\
\hline
\end{tabular}
\caption{LTX constraints on cross-correlations $E_{iso} \propto n_{ref}^w \eta^r$ and $n_{ref} \propto \eta^s$, from equations for $w$, $r$, $s$ provided in the text. Integer entries without errors represent cases where the equation for $s$ is independent of LTX correlation slope $X$. Ranges are provided for $s$ entries in cases where the $s$ equation contains a singularity within the error range of $X$ and all values for $s$ are possible \emph{except} those within the given range. The value $p = 2.2$ has been used for spectral regimes dependent on $p$.}
\label{qrsX_table}
\end{table}

\begin{table}
\centering
\begin{tabular}{rrrrr}
 & & $w$ & $r$ & $s$ \\
\hline
$k = 0$  & D & $-0.06^{+0.02}_{-0.02}$ & $1.39^{+0.14}_{-0.14}$ & $23.82^{+18.18}_{-8.48}$ \\ 
RS  & E & $-0.35^{+0.02}_{-0.02}$ & $1.03^{+0.12}_{-0.13}$ & $2.94^{+0.59}_{-0.54}$ \\ 
 & F & $-0.19^{+0.02}_{-0.03}$ & $0.86^{+0.15}_{-0.15}$ & $4.50^{+1.56}_{-1.25}$ \\ 
 & G & $-0.83^{+0.04}_{-0.04}$ & $-1.68^{+0.20}_{-0.21}$ & $-2.04^{+0.16}_{-0.15}$ \\ 
 & H & $-0.13^{+0.03}_{-0.03}$ & $-0.85^{+0.25}_{-0.26}$ & $-6.38^{+0.45}_{-0.29}$ \\ 
\hline
$k = 0$  & D & $-0.06^{+0.02}_{-0.02}$ & $2.71^{+0.11}_{-0.11}$ & $46.55^{+31.17}_{-14.55}$ \\
FS  & E & $-0.35^{+0.02}_{-0.02}$ & $1.70^{+0.11}_{-0.12}$ & $4.88^{+0.70}_{-0.63}$ \\
 & F & $-0.19^{+0.02}_{-0.03}$ & $1.25^{+0.14}_{-0.15}$ & $6.58^{+1.82}_{-1.46}$ \\
 & G & $-0.83^{+0.04}_{-0.04}$ & $-3.43^{+0.24}_{-0.25}$ & $-4.15^{+0.10}_{-0.10}$ \\
 & H & $-0.13^{+0.03}_{-0.03}$ & $-1.06^{+0.25}_{-0.27}$ & $-8$ \\
\hline
$k = 2$  & D & $-0.20^{+0.08}_{-0.09}$ & $1.31^{+0.18}_{-0.21}$ & $6.61^{+6.06}_{-2.83}$ \\ 
RS  & E & $-3.48^{+0.68}_{-0.98}$ & $-1.22^{+0.79}_{-1.14}$ & $-0.35^{+0.20}_{-0.18}$ \\ 
 & F & $-0.92^{+0.18}_{-0.22}$ & $0.15^{+0.36}_{-0.44}$ & $0.17^{+0.52}_{-0.42}$ \\ 
 & G & $3.80^{+0.30}_{-0.25}$ & $7.66^{+0.40}_{-0.33}$ & $-2.01^{+0.05}_{-0.05}$ \\ 
 & H & $-0.54^{+0.16}_{-0.21}$ & $-1.88^{+0.62}_{-0.78}$ & $-3.46^{+0.15}_{-0.10}$ \\ 
\hline
$k = 2$  & D & $-0.20^{+0.08}_{-0.09}$ & $2.80^{+0.08}_{-0.09}$ & $14.18^{+10.39}_{-4.85}$ \\
FS  & E & $-3.48^{+0.68}_{-0.98}$ & $1.01^{+0.45}_{-0.65}$ & $0.29^{+0.23}_{-0.21}$ \\
 & F & $-0.92^{+0.18}_{-0.22}$ & $2.08^{+0.18}_{-0.22}$ & $2.25^{+0.78}_{-0.63}$ \\
 & G & $3.80^{+0.30}_{-0.25}$ & $10.34^{+0.69}_{-0.56}$ & $-2.72^{+0.03}_{-0.03}$ \\
 & H & $-0.54^{+0.16}_{-0.21}$ & $-2.18^{+0.65}_{-0.82}$ & $-4$ \\
\hline
\end{tabular}
\caption{Same as table \ref{qrsX_table}, now for the LTO correlation.}
\label{qrsO_table}
\end{table}

Although the basic thin shell model does not by itself lead to LTX / LTO correlations, it might still be possible in theory for them to emerge once certain constraints on the underlying physics parameters are met. These constraints could take the form of specific cross-correlations between $E_{iso}$, $n_{ref}$ and $\eta$. In principle, such a cross-correlation between the three progenitor parameters is physically possible and its existence would provide a further constraint on possible progenitor models. More energetic explosions, for example, might be linked to higher Lorentz factor outflows, or be tied to a given type of progenitor star and therefore some regime of associated pre-explosion mass loss that shapes environmental density $n_{ref}$. So, while it should be emphasized that this represents an extension of the basic thin shell model, it is of interest to explore what form such cross-correlations should take in order to satisfy the LTX / LTO correlations. In its most general form, additional cross-correlations are covered by
\begin{eqnarray}
  \label{threepoint_equation}
  E_{iso} & \propto & n_{ref}^w \eta^r, \\
  n_{ref} & \propto & \eta^s,
  \label{twopoint_equation}
\end{eqnarray}
where eq. \ref{threepoint_equation} covers both three-point correlations between all three quantities as well as correlations between energy and density or Lorentz factor separately. These can then be plugged into eq. \ref{T_equation} as well as
\begin{equation}
L \propto E_{iso}^\alpha n_{ref}^\beta \eta^\gamma \propto T^X,
\end{equation}
where coefficients $\alpha$, $\beta$ and $\gamma$ are typical to a given spectral regime and $X$ either the LTX correlation or LTO correlation. In order to obey a given correlation for a given spectral regime, $w$, $r$ and $s$ should obey
\begin{eqnarray}
w & = & (-X - \beta k - 3 \beta) (3 \alpha - \alpha k - X)^{-1}, \nonumber \\
r & = & (-8X + 2 k X - 3 \gamma + \gamma k)(3 \alpha - \alpha k - X)^{-1}, \nonumber \\
s & = & (8X - 2kX + 3 \gamma - \gamma k)(-X - 3 \beta + \beta k)^{-1}.
\end{eqnarray}
For three-point correlations, the conditions on both $w$ and $r$ need to be met simultaneously. In the case of two-point correlations, with only one out of $w$, $r$, or $s$ being applicable, the parameter not affected by the correlation should somehow be universally fixed, in that it either does not vary at all across bursts, or that its range is negligible relative to the impact of the parameter ranges on $T$ and $L_X$ and $L_O$ of the other two physics parameters. In practice, afterglow data analysis efforts since 1997 reveal all three parameters to vary over a substantial range, arguing against the route to LTX / LTO correlations from two-point cross-correlations. The allowed ranges for $w$, $r$, $s$ are tabulated in tables \ref{qrsX_table} and \ref{qrsO_table} for the LTX and LTO correlation respectively.

There exist no possible combinations of spectral regimes that allow for a possible three-point correlation that is able to satisfy both the LTX and LTO correlation, even if a different spectral regime or blast wave region is responsible for LTX than for LTO, either for $k = 0$ or $k = 2$, as can be seen by comparing the permitted $w$ and $r$ intervals from the two tables.

\begin{table}
\centering
\begin{tabular}{rrrrrr}
 & & $E_{iso}$ & $n_{ref}$ & $\eta_{RS}$ & $\eta_{FS}$ \\
\hline
 $k = 0$ & D & - & X & -& -\\
  & E & - & - & -& -\\
  & F & - & - & -& -\\
  & G & - & - & -& -\\
  & H & - & X & X & X \\
\hline
 $k = 2$ & D & - & X & -& -\\
  & E & - & - & -& -\\
  & F & - & - & O & -\\
  & G & - & - & -& -\\
  & H & - & X & X & X \\
\end{tabular}
\caption{Consistency checks if a single variable out of $E_{iso}$, $n_{ref}$, $\eta$ were to dictate the range of both break time $T$ and X-ray / optical luminosity $L$. An X marks consistency with LTX, and O marks consistency with LTO. A value $p=2.2$ has been used.}
\label{single_table}
\end{table}

Another possibility, not requiring additional correlations between $E_{iso}$, $n_{ref}$ and/or $\eta$, would be if not just one, but two out of the three parameters were somehow universally fixed. The correlation would then have to follow completely from the dependencies of time and luminosity on the single varying parameter. In table \ref{single_table}, the options for single parameters to give rise to LTX / LTO correlations are listed. It is not possible to obtain the LTO correlation in the ISM case. In the wind case, the requirement is that the RS emission is dominant, while $\nu_c \ll \nu_m$. Especially in combination with the requirement that not one, but two of the physics parameters have no impact on $T$ and $L_F$, this too does not seem a likely scenario.

In conclusion, neither basic thin shell models nor basic thin shells with three-point correlations between $E_{iso}$, $n_{ref}$, $\eta$ lead to either LTO or LTX correlations. Some possibilities emerge when one of the parameters is taken have no impact on $T$ and $L_X$ / $L_O$, but this seems hard to reconcile with the ranges of parameters that have been reported in the literature (see e.g. \citealt{Soderberg2006, Kocevski2008, Cenko2011, Racusin2011}). When the correlation is obtained through varying a single parameter only, a number of possibilities appear for the $LTX$ correlation, but the $LTO$ correlation remains difficult, and again the price is to assume that parameters actually vary substantially less than reported in the literature.

An alternative possible extension to the basic thin shell model in order to account for LTX and LTO correlations could be the introduction of explicit time dependence in any of the model parameters. In the case of $E_{iso}$, $\eta$ and $n_{ref}$, both break time and luminosity are affected. If this is applied instead to any of the microphysics parameters (i.e., the degree of magnetization $\epsilon_B$, the fraction of energy in shock-accelerated electrons $\epsilon_e$, or the fraction $\xi_N$ of electrons accelerated to a non-thermal distribution), it would affect only the luminosity. Theoretical examples of this approach can be found in \cite{Granot2006, Hascoet2014}. Evidence for evolving microphysical parameters has also been reported in observational studies (e.g. \citealt{Filgas2011, vanderHorst2014}), but these have been inferred for individual cases from the temporal evolution of spectral breaks, rather than consideration of the LTX / LTO correlations. In order to obtain the correlations from evolving parameter observations, it needs to be demonstrated both that the observed time evolution follows the correct power law in time (obeying constraints that can be obtained in a manner similar to those on the internal correlations on the physics parameters) and that the time evolution occurs universally for \emph{all} bursts, rather than individual cases.

\subsection{Thick shells}

For thick shell models with power law energy injection, where $L_{iso} \equiv L_0 t^q$, and $E_{iso} \equiv L_0 T^{q+1} / (q+1)$ and $\eta$ the Lorentz factor of the inflowing wind from the source, the non-correlation between $E_{iso}$ and $T$ implies that $L_0$ and $q$ can not both be universally fixed values. A succesfull progenitor model with a given $q$ (e.g. $q = 0$ for a magnetar, \citealt{Dai1998}) therefore needs to be able to account for the fact that $E_{iso}$ and $T$, rather than $L_0$ and $T$, are independent stochastic variables. In thick shell models with ejecta of mass $M$ accelerated to a single Lorentz factor $\eta$ (also a measure of the Baryon loading of the fireball, according to $E_{iso} / M c^2 = \eta$), the role of energy injection luminosity is played by $\dot{M}$, the mass crossing the reverse shock per unit time.

As was pointed out previously \citep{Leventis2014, vanEerten2014energyinjection}, the LTX and LTO correlations emerge naturally for the thick shell case. For general $k$, we have for the RS region emission:
\begin{eqnarray}
L_D & \propto & E_{iso}^{\frac{15-4k}{3(4-k)}} T^{\frac{2k-9}{3(4-k)}}, \nonumber \\
L_E & \propto & E_{iso}^{\frac{17-6k}{3(4-k)}} T^{\frac{-7}{3(4-k)}}, \nonumber \\
L_F & \propto & E_{iso}^{\frac{16-3k}{4(4-k)}} T^{\frac{5k-16}{4(4-k)}}, \nonumber \\
L_G & \propto & E_{iso}^{\frac{20-5k-pk}{4(4-k)}} T^{\frac{3k-12-pk}{4(4-k)}}, \nonumber \\
L_H & \propto & E_{iso}^{\frac{16-2k-pk}{4(4-k)}} T^{\frac{6k-16-pk}{4(4-k)}}, \nonumber
\end{eqnarray}
where we kept $E_{iso}$ explicit as well, given its potential to skew the LTX / LTO correlation via an implicit dependence on $T$, as discussed above. For the FS region emission, we have
\begin{eqnarray}
L_D & \propto & E_{iso}^{\frac{10-4k}{3(4-k)}} T^{\frac{2-k}{4-k}}, \nonumber \\
L_E & \propto & E_{iso}^{\frac{14-6k}{3(4-k)}} T^{\frac{2-3k}{3(4-k)}}, \nonumber \\
L_F & \propto & E_{iso}^{\frac{3}{4}} T^{\frac{-1}{4}}, \nonumber \\
L_G & \propto & E_{iso}^{\frac{12-5k+4p-pk}{4(4-k)}} T^{\frac{12-5k-12p+3pk}{4(4-k)}}, \nonumber \\
L_H & \propto & E_{iso}^{\frac{2+p}{4}} T^{\frac{2-3p}{4}}, \nonumber
\end{eqnarray}
Specific values for $k =0, 2$ and $p = 2.2$ are included in table 4 of \cite{vanEerten2014energyinjection}. The synchrotron slope $p$ typically lies around $\sim 2.2$ (either at a universal value, observed within a range due to measurement errors, or intrinsically distributed across some range), both according to theory (e.g. \citealt{Kirk2000, Achterberg2001}) and observations (e.g. \citealt{Curran2009, Ryan2014}). As discussed in \cite{vanEerten2014energyinjection}, the thick shell equations above imply that, for the cases $k = 0$ (``interstellar medium'', ``ISM'') and $k = 2$ (``wind''), there is no clear support for the LTO correlation from FS emission (and multiple options from RS emission), although $L_G \propto T^{-(0.9 + 0.75 \Delta p)}$, where $\Delta p \equiv p - 2.2$, comes closest to achieving the LTO correlation from FS emission. Furthermore, although the error bars in the reported correlation are too large to admit definitive statements, $L_H$ generally comes closest to explaining the LTX correlation. For FS emission we have $L_H \propto T^{-(1.15+0.75\Delta p)}$, independent of $k$. For RS emission in a wind, we have $L_H \propto T^{-(1.05 + 0.25 \Delta p)}$.

\section{Correlations in a synthetic light curve population}
\label{synthetic_section}

\begin{table*}
\centering
\begin{tabular}{rrrrrrrrrr}
\hline
& & thin & thin & thin & thin & thick & thick & thick & thick \\
& & FS & RS & FS & RS & FS & RS & FS & RS \\
& & ISM & ISM & wind & wind & ISM & ISM & wind & wind \\
\hline
$\left<m\right>$ & LTX & 1.6 & 2.0 & 4.9 & 6.3 & 0.6 & 0.8 & 1.9 & 0.7 \\
 & LTO & 0.8 & 2.2 & 10. & 12. & 0.9 & 0.8 & 0.7 & 0.9 \\
$ \left< \rho \right>$ & LTX & $10^{-41}$ & $10^{-31}$ & $10^{-143}$ & $10^{-132}$ & $10^{-5}$ & $10^{-4}$ & $10^{-6}$ & $10^{-4}$ \\
& LTO & $10^{-25}$ & $10^{-24}$ & $10^{-88}$ & $10^{-89}$ & $10^{-3}$ & $10^{-2}$ & $10^{-5}$ & $10^{-3}$ \\
& $E-T$ & $10^{-6}$ & $10^{-6}$ & $10^{-6}$ & $10^{-6}$ & 0.5 & 0.5 & 0.5 & 0.5 \\
\hline
\end{tabular}
\caption{First part: average consistency measures $\left< m \right>$ between LTX / LTO correlations and 10,000 runs of 500 bursts each, for different environments, emission region and shell types. When $\left< m \right> < 1$, the synthetic samples are consistent with the correlation reported in the literature within their $1 \sigma$ error bars. Second part: average probability (over 10,000 runs) of chance correlation $\left< \rho \right>$ within a synthetic sample, between $L_X$ and $T$, $L_O$ and $T$, $E_{iso}$ and $T$ from top to bottom. The exact threshold for declaring two parameters to be correlated is arbitrary, but will have $\left< \rho \right> \ll 1$. In this table, this condition is met everywhere, except for the thick shell correlations between $E_{iso}$ and $T$.}
\label{population_table}
\end{table*}

One can test what correlations emerge in practice by sampling a set of artificially generated light curves based on some underlying model (e.g. thick or thin shell) and assumed underlying populations for the various model parameters. Unfortunately, the real distributions occurring in nature for such parameters as $E_{iso}$, $n_{ref}$ and $T$ are not well known, although we can make some broad estimates based on the accumulated results of afterglow analysis so far. For this reason, one should not draw strong conclusions if they depend sensitively on the shape of any of the model parameter distributions and we will avoid doing so. Instead, we use the analysis below to make two points that turn out to be robust under changes in the underlying distribution: (1) it remains difficult to reconcile the thin shell model with the observed correlations and (2) the thick shell model preference for RS emission in order to explain the LTO correlation is no longer significant when a sample of artificial bursts is considered.

A large number (10,000) of collections of break times and fluxes, 500 bursts each, are generated randomly, both for thick and thin shell models, based on flux equations from \cite{vanEerten2014energyinjection} and \cite{Yi2013}, respectively. The spread of the inferred correlation slopes increases with decreasing sample size, going to e.g. 50 bursts each, as one would expect, and decreases for increasing sample size (e.g. 5,000). However, this is found to not impact the results. The illustrative value of 500 reflects the order of the size of the Swift XRT sample. Unless otherwise specified, the following model parameter values  are used. Redshift is kept fixed at $z=0$, $d_L$ set to $10^{28}$ cm; no attempt will be made to capture the observed scale factor in front of the correlations (i.e. $\log C$, from $\log L = \log C + \alpha \log T$). $\log E_{iso}$ is drawn from a Gaussian distribution centered at 53 (i.e. $E_{iso} = 10^{53}$ erg), with standard deviation $\sigma = 1$. For the ISM case, $\log n_{ref}$ is drawn from a Gaussian distribution centered at $\log 1$, with standard deviation $\sigma = 2$, for the wind case $\log n_{ref} = \log 29.89$ (at $R_{ref} = 10^{17}$ cm), with the same standard deviation. These energy and density ranges were informed by \cite{Cenko2011}. A Synchrotron slope $p$ is drawn from a Gaussian distribution peaking at 2.2, with $\sigma = 0.1$ and values $p < 2.01$ are redrawn in order to avoid the need to introduce a more complicated synchrotron model with upper cut-off Lorentz factor for the accelerated particle population. As discussed previously, these $p$ values are consistent with \cite{Kirk2000, Achterberg2001, Curran2009, Ryan2014}. Accelerated electron energy fraction $\epsilon_e$ is kept at 0.1, accelerated electron number density fraction $\xi_N$ at 1, and magnetic field enery fraction $\epsilon_B = 10^{-2}$. For thin shells, ejecta Lorentz factor $\log \eta$ is drawn from a Gaussian distribution centered at $\log 25$, with $\sigma = 0.5$. For thick shells, $\log \eta$ peaks at $\log 10^4$, with the same standard deviation. Note that, in the thick shell case, $\eta$ is the Lorentz factor of the material entering the ejecta through the RS, and the Lorentz factor of the shocked ejecta itself can be much smaller (see e.g. \citealt{vanEerten2014energyinjection}). For the thick shell case, energy injection durations $T$ are drawn from a Gaussian distribution in log space with $\log T$ peaking at $\log 5 \times 10^3$ and $\sigma = 0.5$. For thin shells, $T$ is calculated according to eq. \ref{T_equation}, with pre-factor given by \cite{Yi2013}.

\subsection{Thin shells}

\begin{figure*}
 \centering
  \includegraphics[width=\columnwidth]{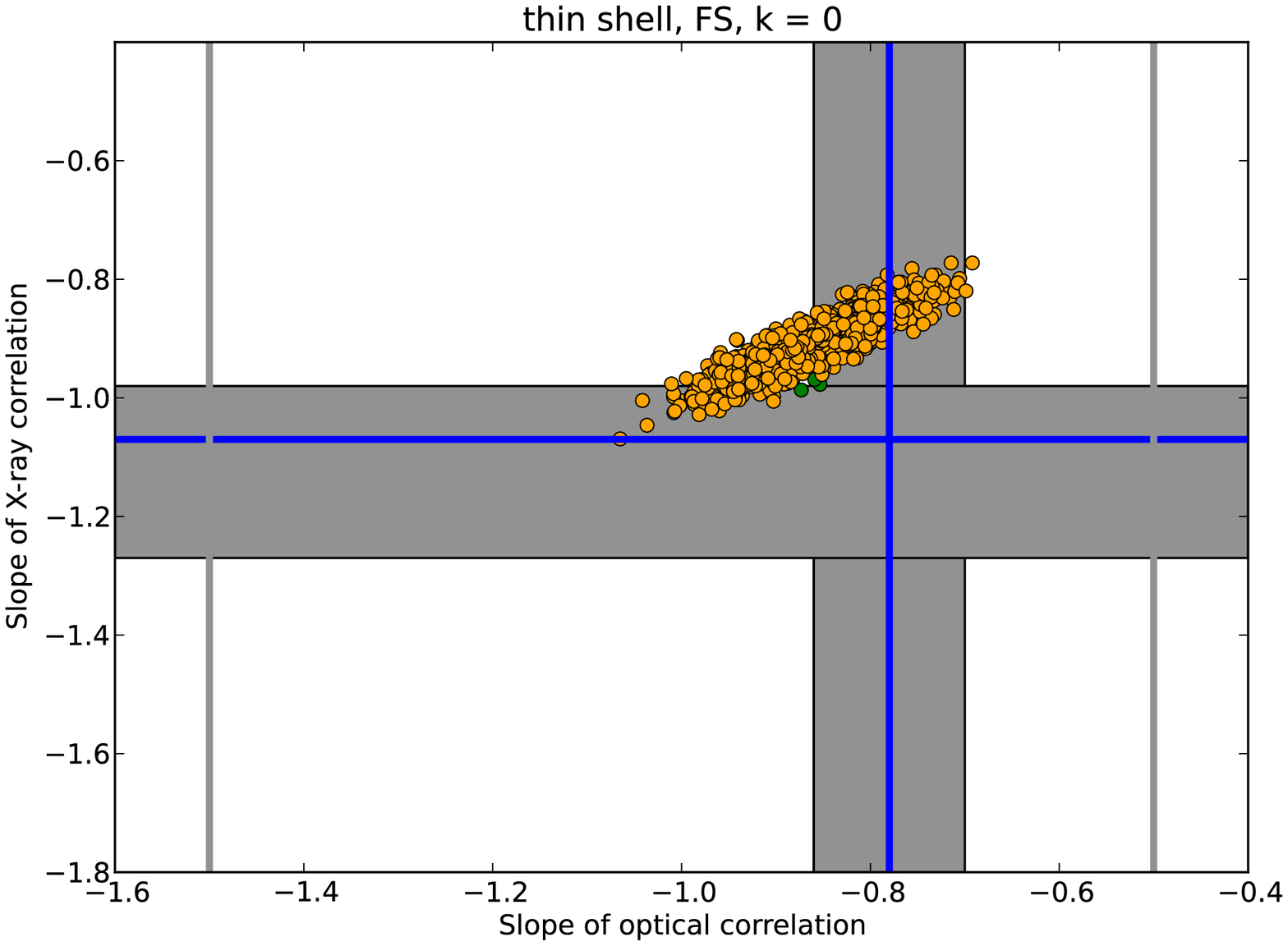}
  \includegraphics[width=\columnwidth]{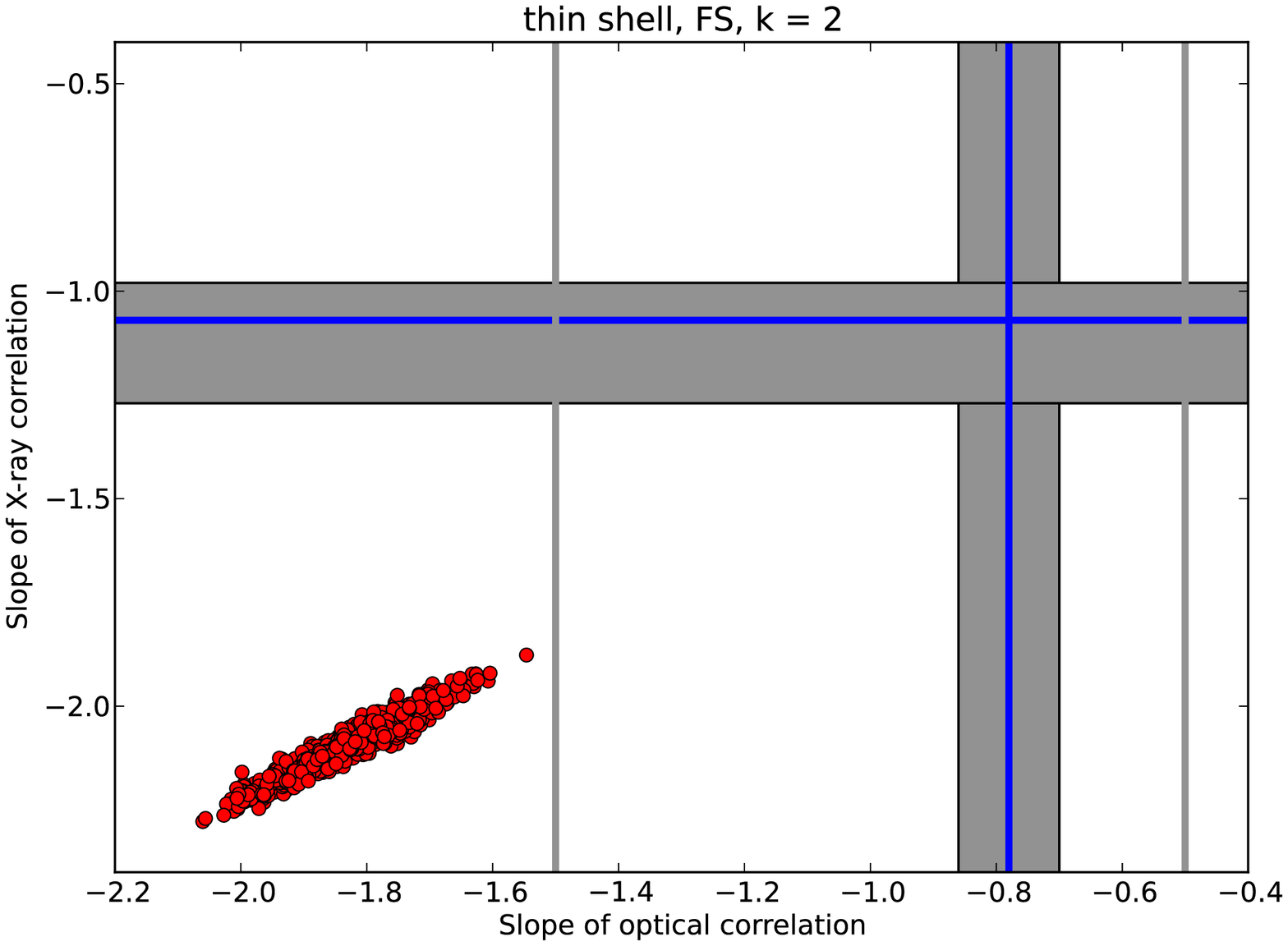}
  \includegraphics[width=\columnwidth]{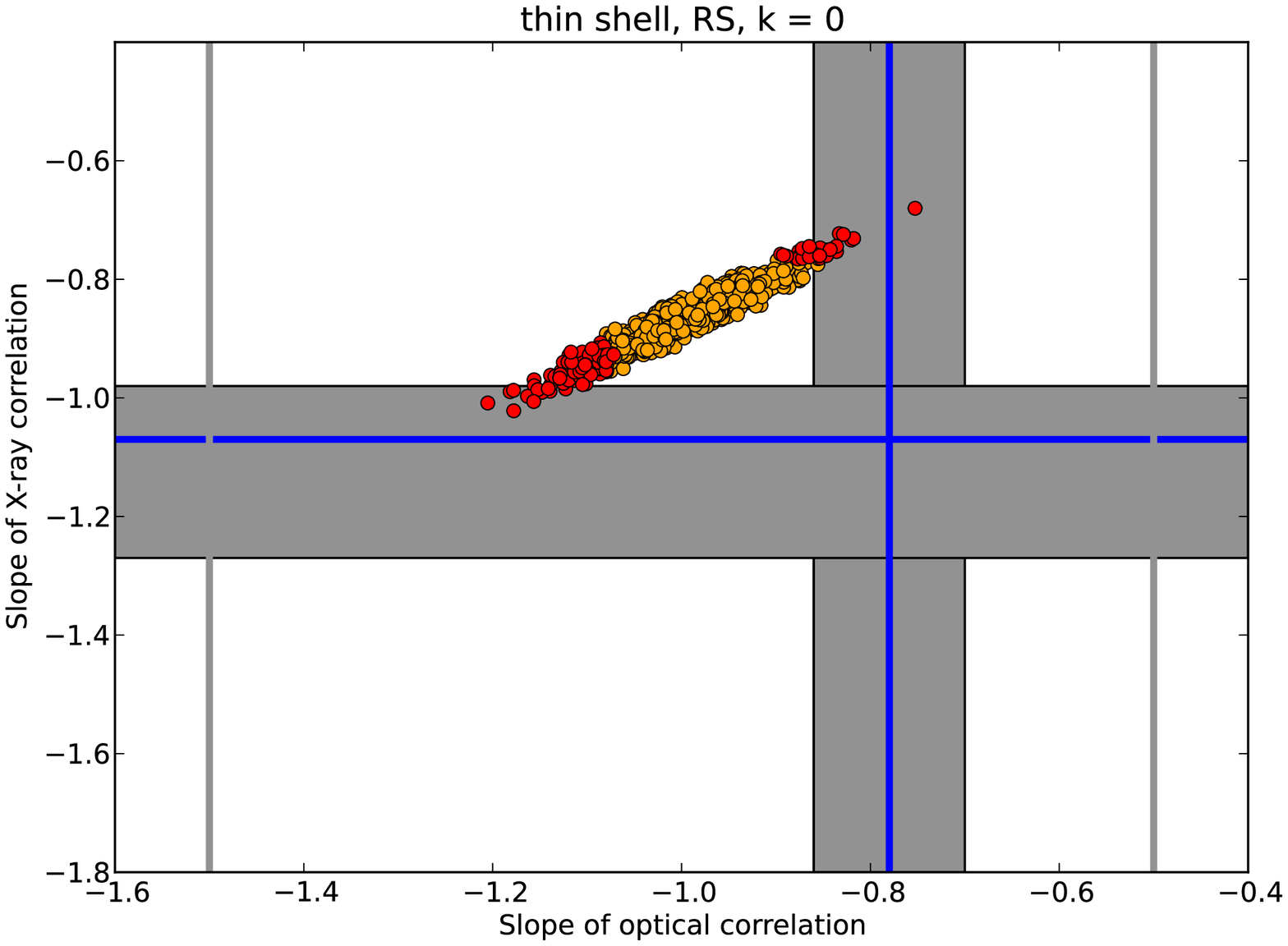}
  \includegraphics[width=\columnwidth]{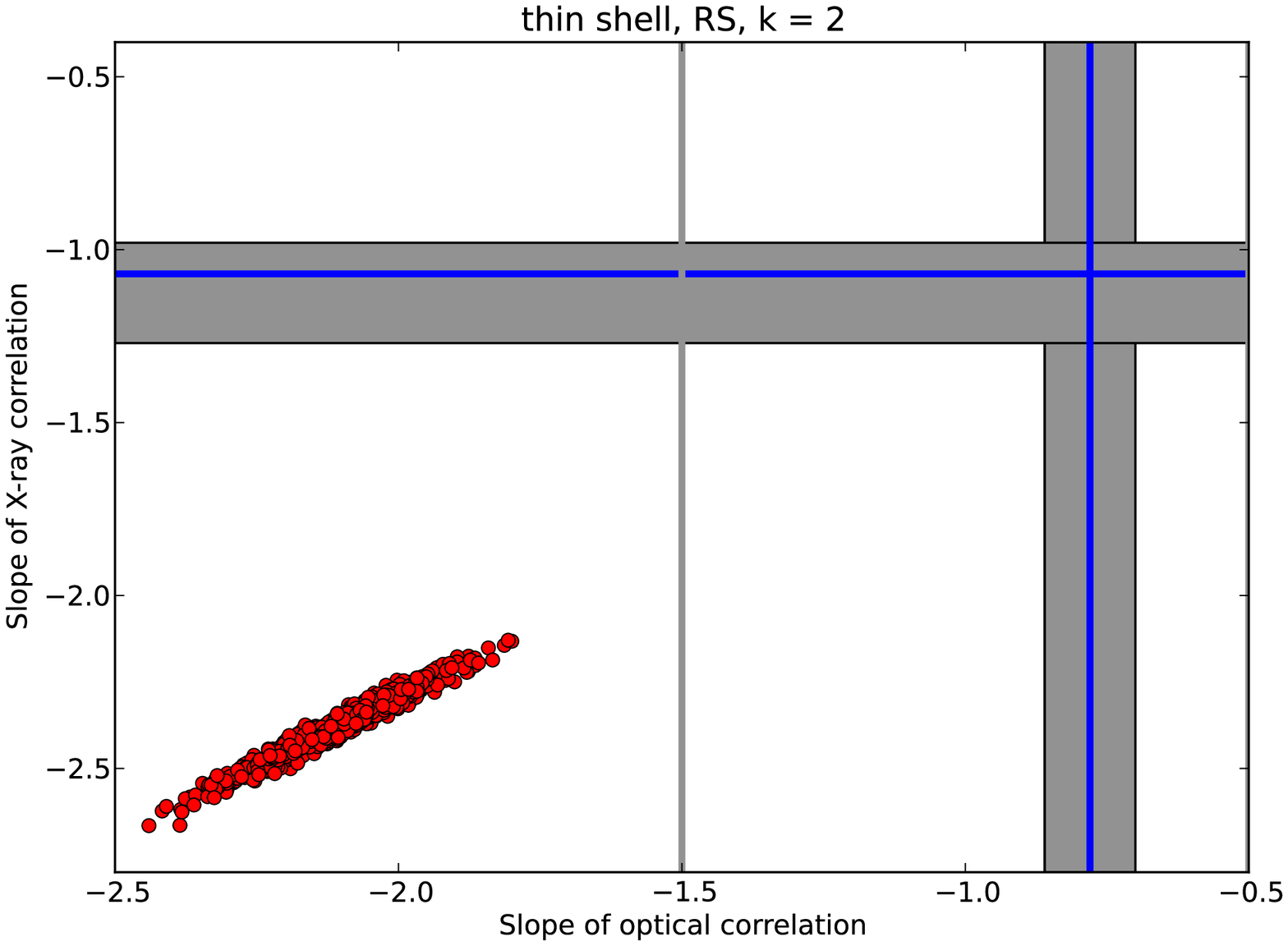}
\caption{Comparison between correlation slopes for 1000 thin shell sample runs and observational LTO (horizontal direction) and LTX (vertical direction) slopes. Grey band indicate 1 $\sigma$ errors on LTX / LTO correlations. Green dots are runs consistent within 1$\sigma$ error bars for both (according to eq. \ref{consistency_equation}), orange dots are consistent within 3$\sigma$ for both but not 1$\sigma$ for both and red dots pass neither test. Vertical grey lines denote wider LTO error bars from \protect\cite{PanaitescuVestrand2011}.}
 \label{alpha_thin_figure}
\end{figure*}

For all of the four permutations of (RS, FS) and $k = (0, 2)$, the correlation between $\log E_{iso}$ and $\log T$ is found to remain strongly intact: a Spearman rank test indicates correlations with chance probabilities $\rho \ll 1$ for all individual runs, as tabulated in table \ref{population_table}.

Searches for LTX / LTO type correlations for thin shell models also reveal very small chance probabilities, implying that correlations between these quantities genuinely emerge from the samples (see table). However, as anticipated, this emergence is driven by the shape of the underlying model parameter distributions. If, for example, the width of the $\log n_{ref}$ distribution is changed from 2 to 1 (such that it becomes comparable to the range of e.g. $E_{iso}$), the existence of correlations becomes doubtful in the ISM case and $\left< \rho \right> \sim 0.3$ for both LTX, LTO in both FS and RS cases (but not the existence of a correlation between $E_{iso}$ and $T$, which remains intact). 

Regardless, the best fits for $\log L = \log C + \alpha \log T$ are inconsistent with the LTX and LTO correlation $\alpha$ values. A consistency test using the measure
\begin{equation}
m = \frac{| \alpha_{observed} - \alpha_{synthetic} |}{\sqrt{\sigma^2_{observed} + \sigma^2_{synthetic}}},
\label{consistency_equation}
\end{equation}
where (in the asymmetric LTO case) $\sigma_{observed}$ the error in the direction of $\alpha_{synthetic}$, generally results in (on average for 10,000 runs) $\left< m \right> > 1$, as shown in the table. An $m$ value $<1$ is expected for full consistency within $1 \sigma$ error bars between synthetically generated and observed LTX / LTO correlations, a value $m < 3$ indicating that the synthetic result comes interestingly close. This is also illustrated visually in figure \ref{alpha_thin_figure}. Here the best fit correlation slopes for a subset of 1000 samples are overplotted on the LTX / LTO correlations from the literature. Because each individual correlation measure from a synthetic sample has an associated error bar, the data points do not need to lie exactly within the crossing of the literature correlation bands in order to be consistent. This is indicated with the color coding of the data points (the thick shell analogue of this plot, provided by Fig. \ref{alpha_figure} and discussed below, for example, shows some green data points outside of the square where the two bands overlap).

Altogether, the population study demonstrates that it remains extremely difficult to reconcile the basic thin shell model with the LTX / LTO correlations and the expected $E_{iso} - T$ non-correlation. In reality, even thin shells will not have a sharply defined back as assumed by simplified crossing time calculations. However, if one tries to use this aspect to weaken the $E_{iso} - T$ correlations, one also further weakens the LTX / LTO correlations by the same token. It also makes it even more difficult to explain values of the reported LTX / LTO correlations.

\subsection{Thick shells}

\begin{figure*}
 \centering
  \includegraphics[width=\columnwidth]{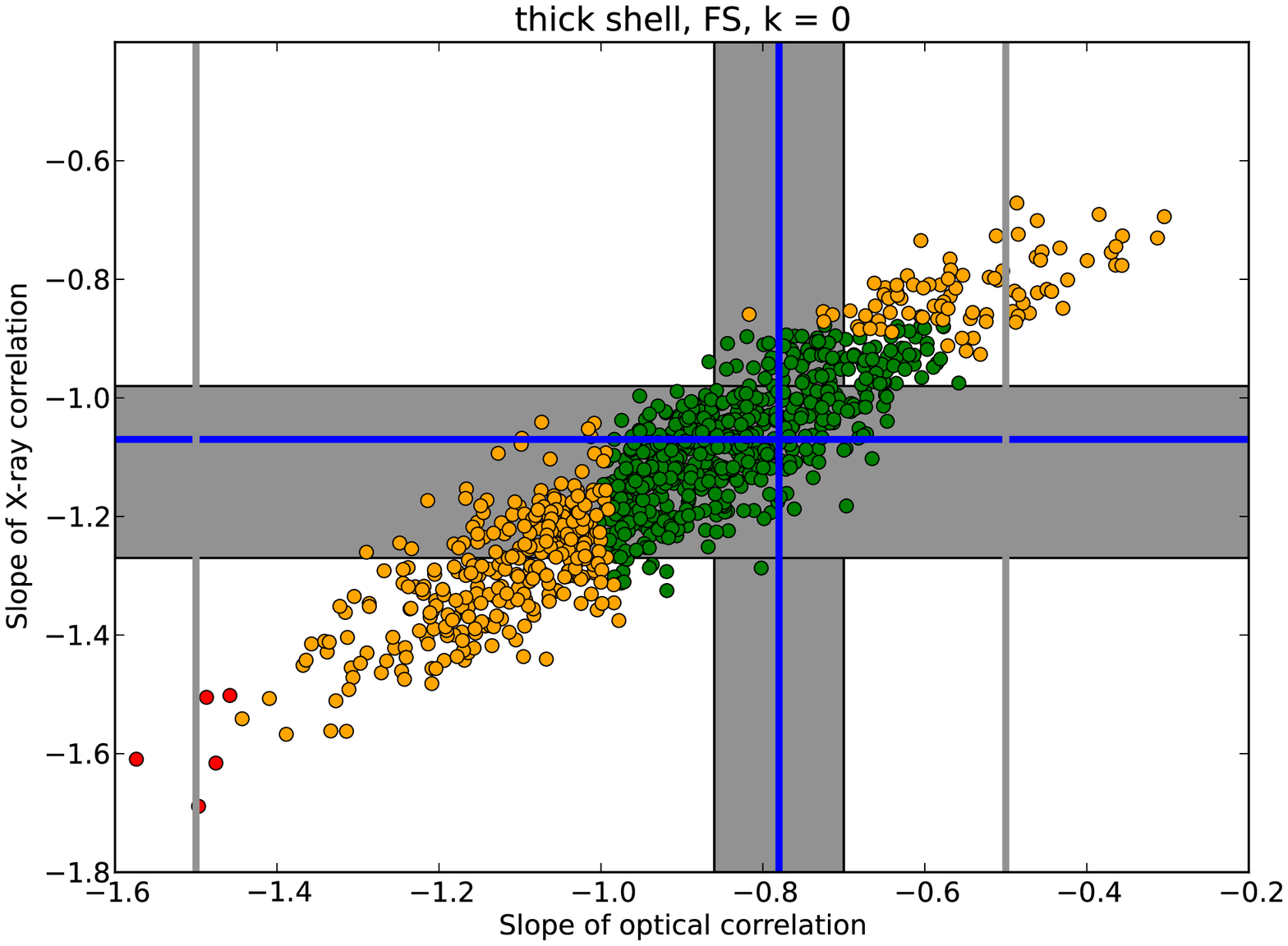}
  \includegraphics[width=\columnwidth]{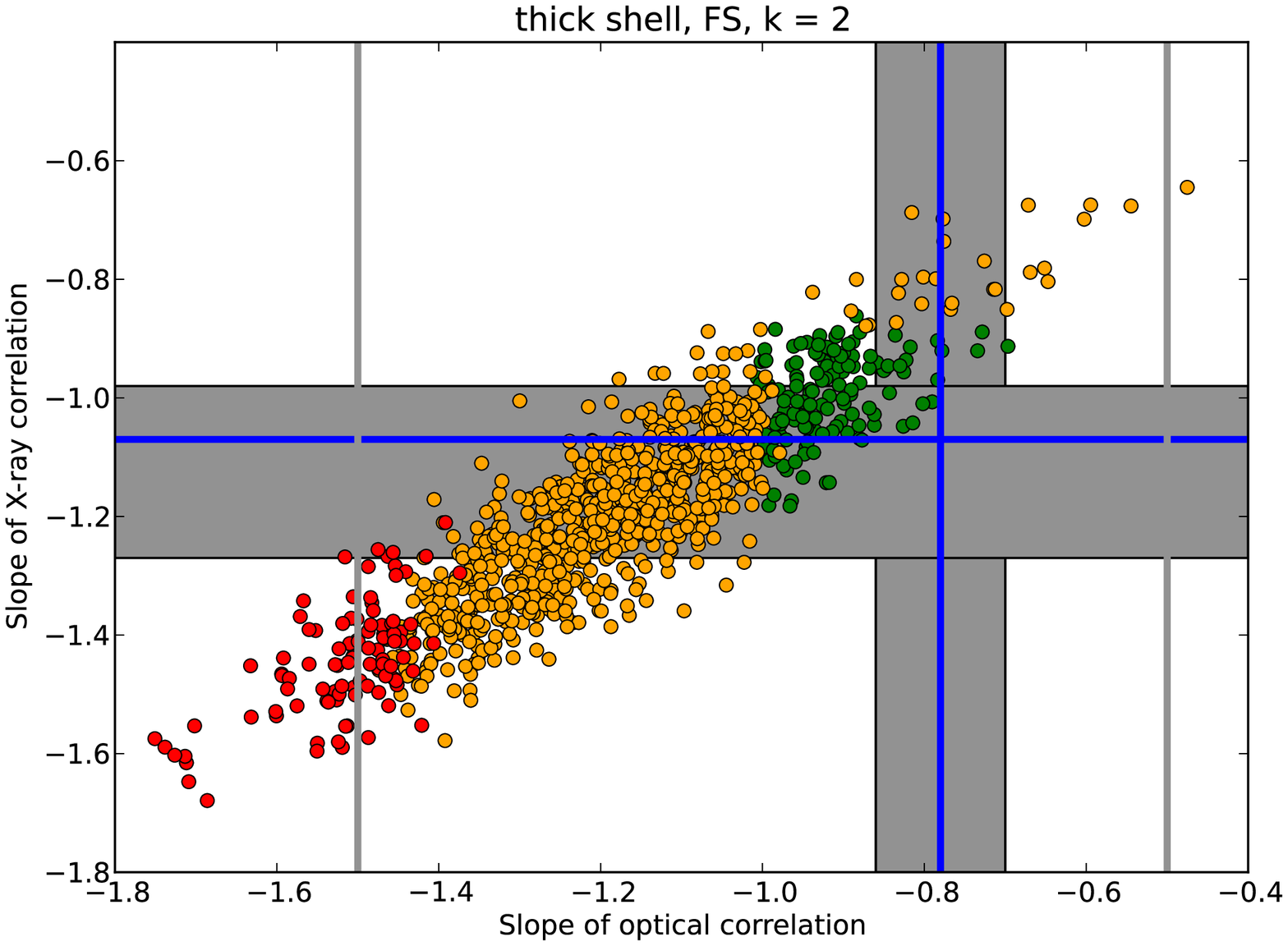}
  \includegraphics[width=\columnwidth]{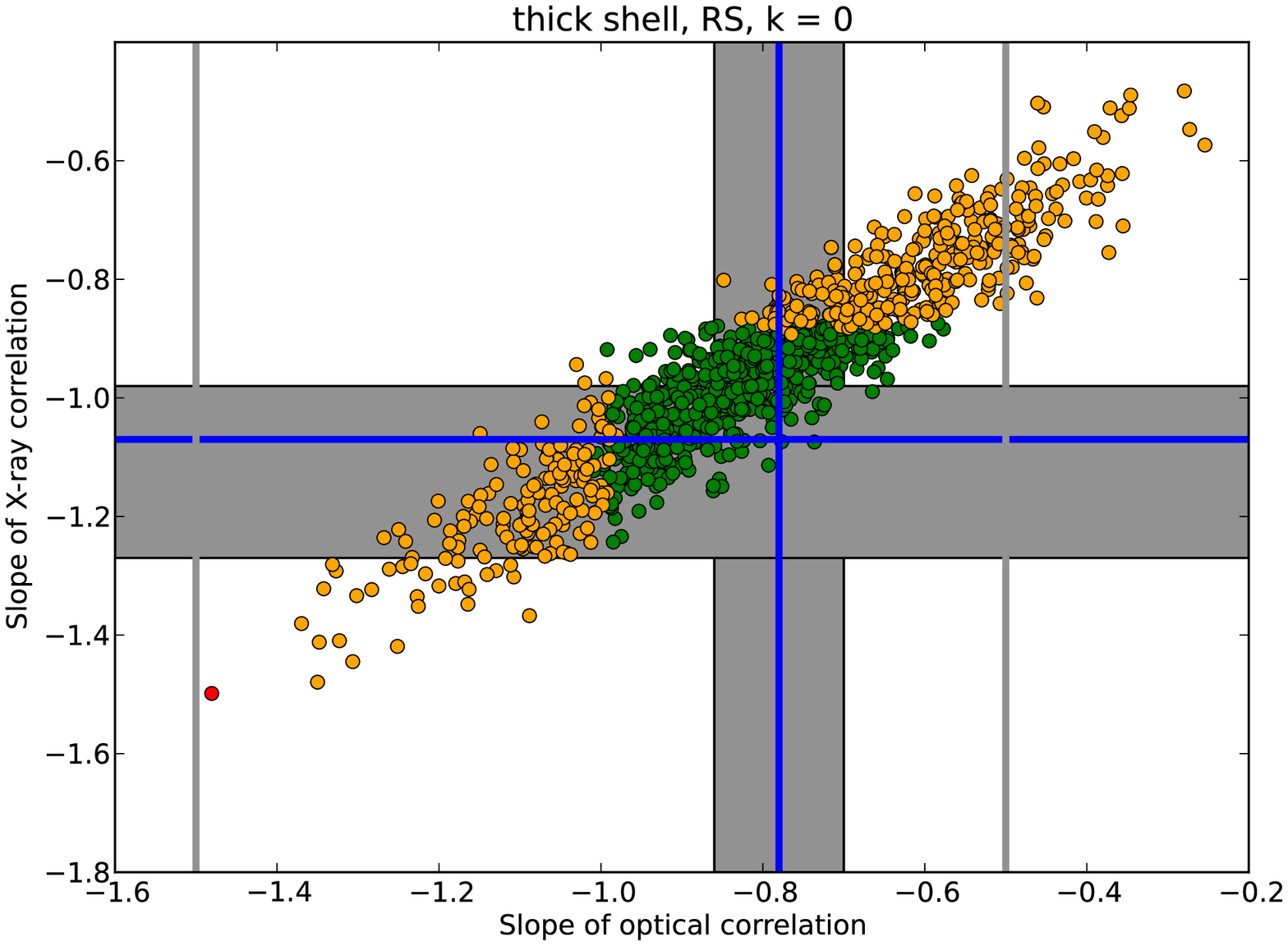}
  \includegraphics[width=\columnwidth]{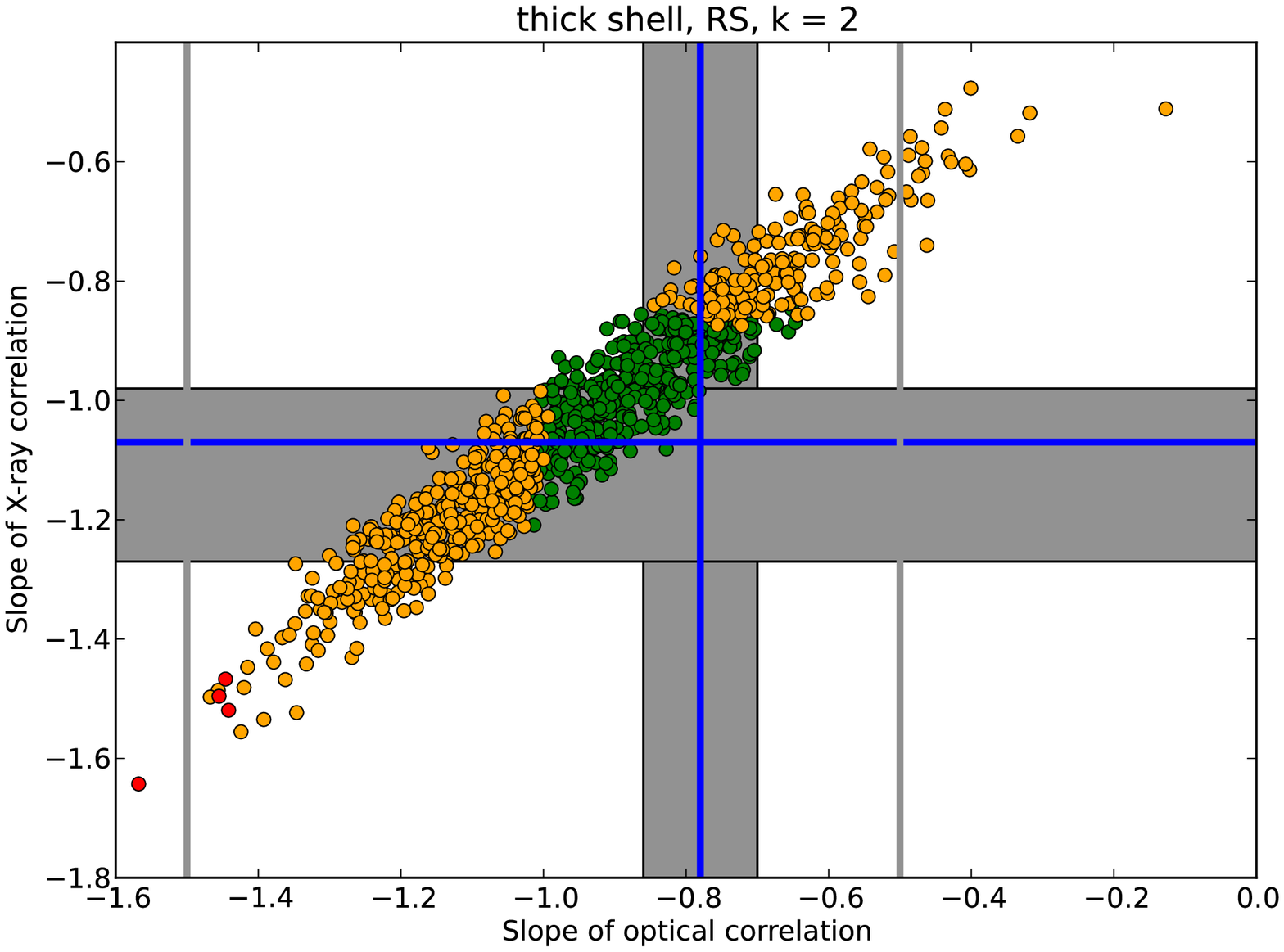}
\caption{Same as Fig. \ref{alpha_thin_figure}, now for thick shells: comparison between correlation slopes for 1000 sample runs and observational LTO (horizontal direction) and LTX (vertical direction) slopes. Grey band indicate $1 \sigma$ errors on LTX / LTO correlations. Green dots are runs consistent within 1$\sigma$ error bars for both (according to eq. \ref{consistency_equation}), orange dots are consistent within 3$\sigma$ for both but not 1$\sigma$ for both and red dots pass neither test. Vertical grey lines denote wider LTO error bars from \protect\cite{PanaitescuVestrand2011}.}
 \label{alpha_figure}
\end{figure*}

\begin{figure*}
 \centering
  \includegraphics[width=\columnwidth]{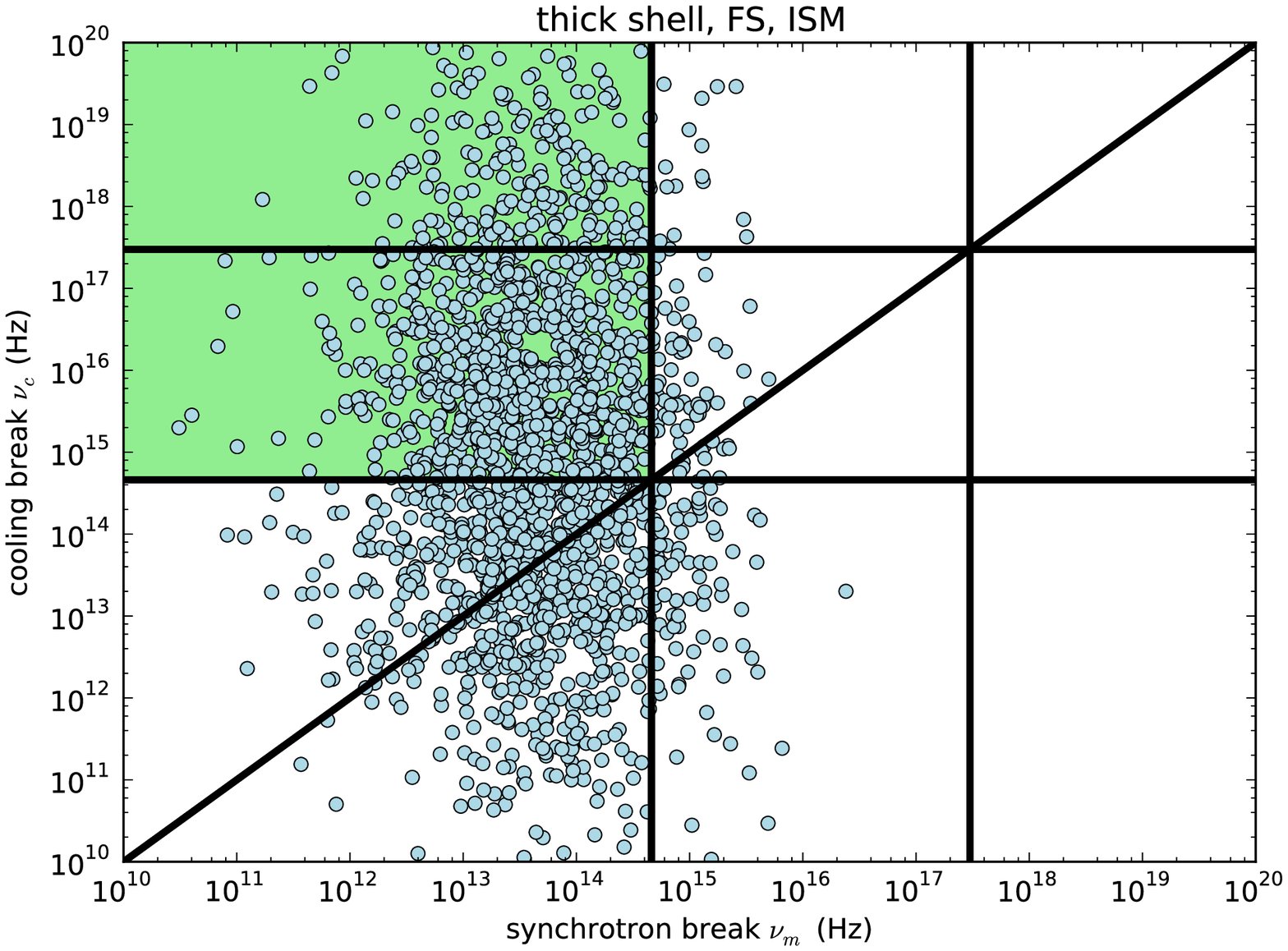}
  \includegraphics[width=\columnwidth]{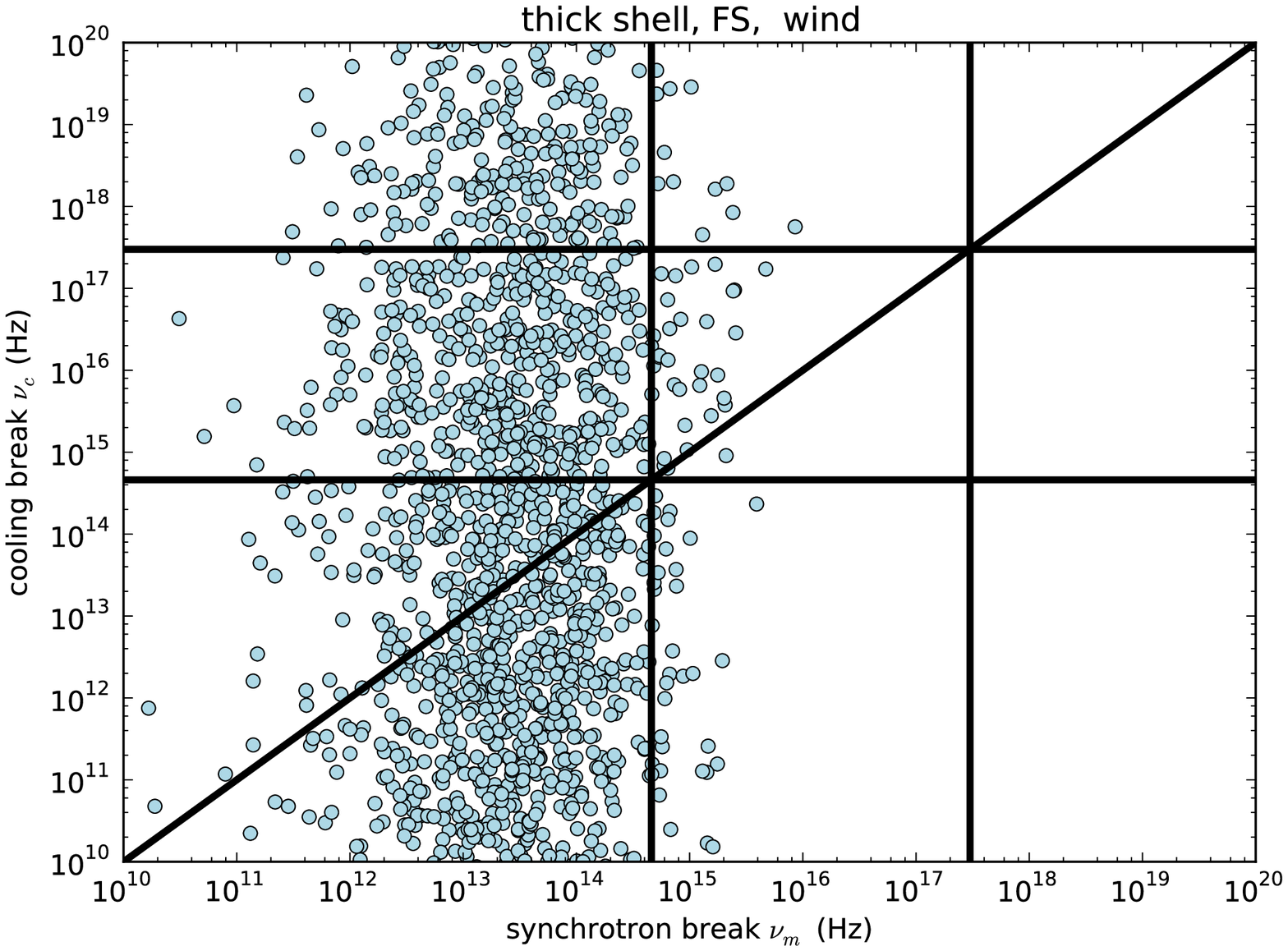}
  \includegraphics[width=\columnwidth]{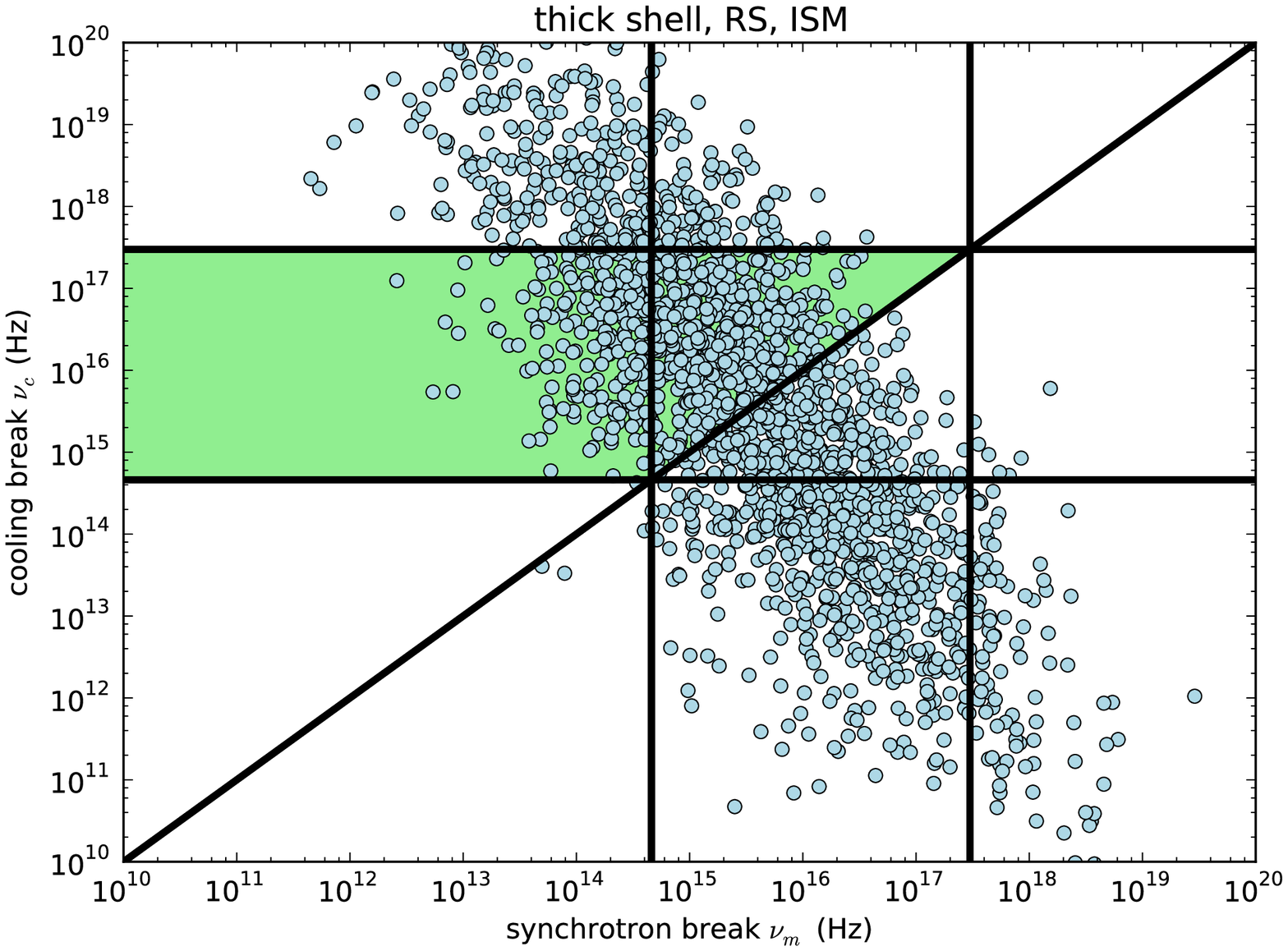}
  \includegraphics[width=\columnwidth]{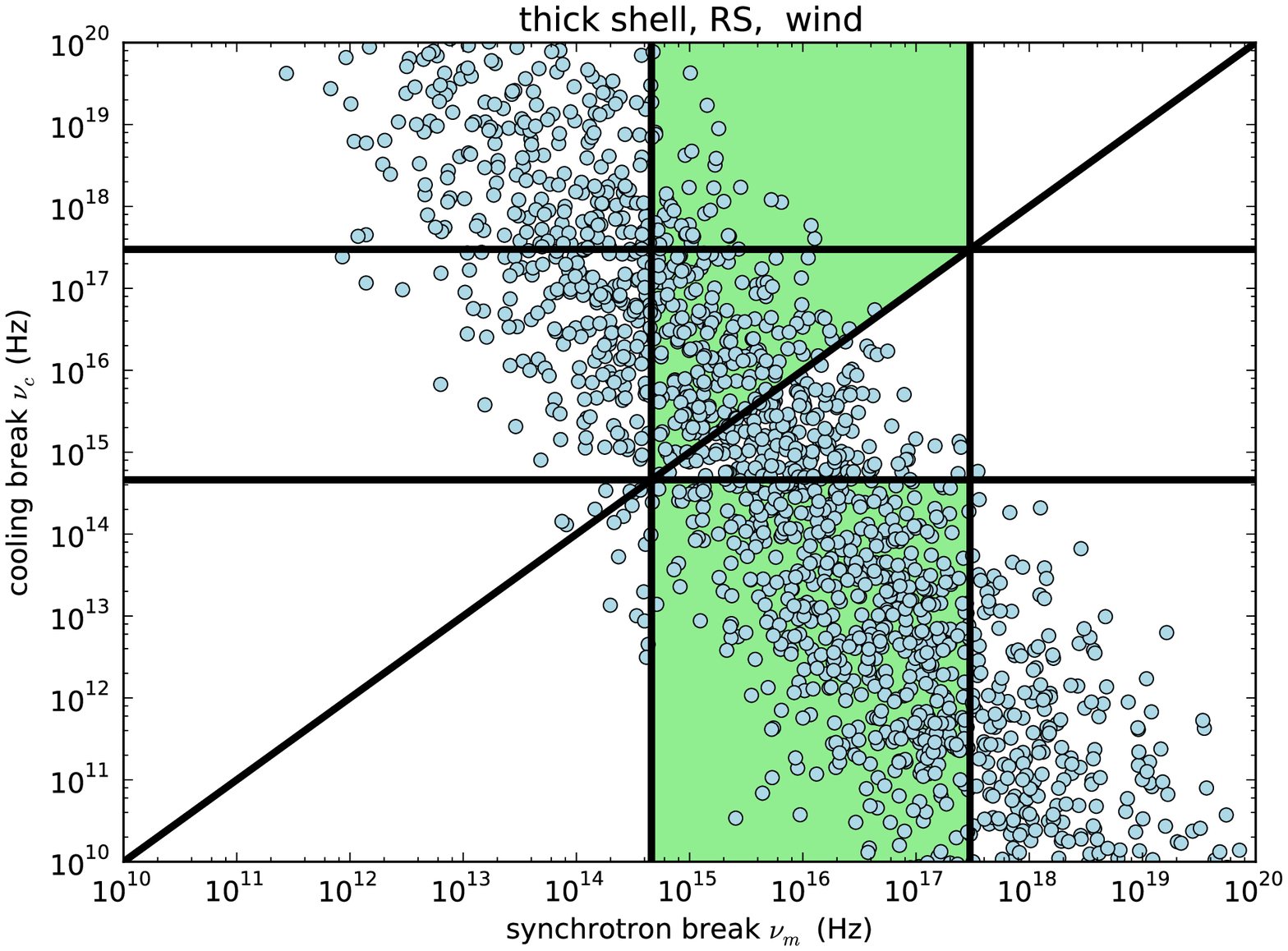}
\caption{Characteristic frequencies for single samples of 2000 bursts in the thick shell scenario. Each dot represents a single combination of $\nu_m$ (horizontally) and $\nu_c$ (vertically). Observer frequencies at X-rays ($3 \times 10^{17}$ Hz) and R-band ($4.6 \times 10^{14}$ Hz) are overplotted with horizontal and vertical lines. The diagional line marks the point where $\nu_m = \nu_c$. Green colored regions of parameter space are consistent with both the LTX and LTO correlations (this depends on in which spectral regimes optical and X-ray observations fall for a given region).}
 \label{numnuc_figure}
\end{figure*}

When samples are generated for the thick shell scenario, no spurious correlations between $E_{iso}$ and $T$ are found on average, with $\rho = 0.5$ for all four (RS, FS) and $k = (0, 2)$ permutations. The average chance probabilities for all correlations are reported next to the thin shell case in table \ref{population_table}. The weakest reported correlation is for LTO, RS, $k = 0$.

The match between the reported correlations from the sample runs and the reported LTX / LTO correlations are shown in Fig. \ref{alpha_figure}. The main point from these figures is that, on the whole, there are various routes to reproducing the LTX / LTO correlations even when using the same emission region (RS or FS) to account for both. For the underlying model parameter distributions that have been used, FS emission from thick shells in a stellar wind environment performs the poorest in reproducing the LTX correlation. 

\begin{figure}
 \centering
  \includegraphics[width=\columnwidth]{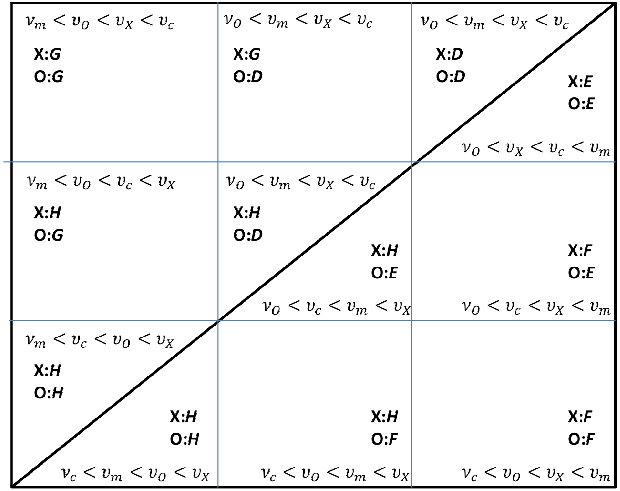}
\caption{Overview of possible ordering of observation frequencies in optical ($\nu_O$) and X-ray ($\nu_C$), and characteristic frequencies $\nu_m$ (synchrotron injection break) and $\nu_c$ (synchrotron cooling break), for comparison with the burst populations plotted in fig. \ref{numnuc_figure}. Again, the diagonal line marks where $\nu_m = \nu_c$. The left vertical line / lower horizontal line marks the optical observation frequency, the right vertical line / upper horizontal line marks the X-ray frequency.}
 \label{tiles_figure}
\end{figure}

More insight into how exactly the correlations follow from the thick shell model luminosity equations, can be gained from examining a single sample in detail. In Fig. \ref{numnuc_figure}, the combinations of characteristic frequencies $\nu_m$ and $\nu_c$ are plotted for each burst in single a synthetic sample. The parameter space is divided into a number of regions by the X-ray and optical observer frequencies and along the line where $\nu_m = \nu_c$. Each tile in Fig \ref{numnuc_figure} represents a certain combination of spectral regimes for optical and X-ray observations, according to eq. \ref{spectral_regime_equations}. The leftmost tile on the middle row corresponds to X-ray observations in regime $H$ and optical observations in regime $G$, etc. In Fig \ref{tiles_figure}, all possible spectral orderings are listed, as well as the spectral regimes in optical (O) and X-rays (X). Tiles where the model predictions are consistent with the LTX / LTO correlations are colored green in Fig. \ref{numnuc_figure} (no account was made for the fact that optical and X-ray observations typically probe different spectral regimes, as discussed in section \ref{key_correlations_section}. Using Fig. \ref{tiles_figure}, one can identify where the optical and X-ray spectral regimes are identical. In principle, this could be used in a future population study to constrain the possible underlying model parameter distributions). The consistency between a luminosity equation and the LTX / LTO correlations  is determined assuming $p$ to lie within the range $2.07 - 2.51$ (see \citealt{vanEerten2014energyinjection, Ryan2014}). Note, however, that the correlation slopes from the synthetic sample are calculated from the entire sample. It is therefore not necessary for all individual bursts to lie within green tiles. In fact, given the error bars on the correlation slopes calculated from the sample (with sizes determined by the ranges of the underlying distributions of the model parameters and the number of burst per sample), it is sometimes even possible to reproduce a slope consistent with the literature without any individual burst in the sample within a green tile (compare e.g. the FS wind case without green tiles to the slope plot from Fig. \ref{alpha_figure}, which nevertheless does show a number of green data points).

The implication is that thick shells are capable of reproducing the observed LTX / LTO correlations without requiring overly specific underlying parameter distributions.

\section{Discussion and summary}
\label{discussion_section}

The existence of various correlations between parameters (e.g. luminosity, characteristic break times, fluence) describing GRB prompt emission and afterglow light curves at early and late times, is a striking result that emerges whenever large samples of GRBs are studied. Ideally, these correlations can be used to test model predictions (as done in e.g. \citealt{DadoDar2013}, for ``cannonball'' type models) and distinguish between models capable of reproducing the correlations and those that require either fine-tuning or are falsified altogether.

One way of obtaining specific correlations is via introducing time dependency in the parameters describing the microphysics of the radiation (see e.g. \citealt{Granot2006, Hascoet2014} for examples involving $\epsilon_B$, $\epsilon_e$), but this additionally requires an underlying microphysical model justifying the precise nature of the newly introduced physics (e.g. why $\epsilon_e$ depends on circumburst density, not blast wave Lorentz factor, again see \citealt{Hascoet2014}).

Here I take a more limited approach and stay with the standard non-changing microphysics assumption for relativistic blast waves in the context of thick and thin shell models. In the thin shell model, the afterglow plateau phase is the result of the pre-deceleration emission from a slower component in a two-component or jet-cocoon type model. For thick shells, the plateaus result from energy injection either in the form of late activity from the source of via additional kinetic energy from slower ejecta catching up with the blast wave, as long as the amount of energy injected remains sufficiently large to allow for a relativistic reverse shock.

It is shown that thin shell models can not be reconciled with the observed LTX / LTO correlations between afterglow plateau end time and luminosity and that they imply the existence of a correlation between plateau end time and ejecta energy that is not seen in the data. Basic thin shell models where the underlying physics parameters, explosion energy, circumburst density and ejecta Lorentz factor, remain uncorrelated do not lead to a correlation between time and luminosity, while no three-point correlations between the three physics parameters are possible that can explain both the LTX and LTO correlation even when each of those is shaped by a different dominant emission region or spectral regime. This does not mean that successful data fits using a thin shell model are not possible. In theory, it might even be possible to successfully fit all bursts with plateau stages in this way. However, this study demonstrates that such an effort will inevitably lead to a sample whose properties as a whole can not be explained from the basic thin shell model alone, and within which additional model parameter correlations will have emerged.

Thick shell models, on the other hand, can easily reproduce the LTX / LTO correlations across a range of uncorrelated underlying values for the model parameters. They do this so well, in fact, that it is unfortunately difficult to distinguish in this way between forward shock (FS) and reverse shock (RS) emission dominated models, or homogeneous and stellar wind-type environments. By definition, the observed flux is shaped by simultaneous emission from both regions. Whether one region dominates or whether the two contributions are comparable, depends on the values for the model parameters (and possibly on differences in their microphysics parameters, such as $\epsilon_B$). In the case of comparable contributions, the existence of a clear LTX correlation implies that both regions emit in one of the allowed spectral regimes, although not necessarily in the same one.

It is tempting to take the falsification of the basic thin shell model in its simple form as an argument against the collapsar nature for GRBs with plateaus in their afterglows, since the traditional single thin shell and two-component jet-cocoon system cannot explain, respectively, the existence of plateaus and the luminosity - time correlations involving plateaus. However, this is likely an overintepretation of an overly simplified model, and collapsar outflows probably involve a range of Lorentz factors that thereby can account for late energy injection into the forward shock, moving the collapsar afterglow predictions into thick shell territory. In any case, the results from this study are certainly consistent with long term energy injection, as expected e.g. from a magnetar model. Thick shell models are capable of reproducing the LTX / LTO correlations independent of the value of $q$, which drops out of the equations at when observing at $T$.

An interesting possibility for further study is the potential for FS and RS regions to jointly shape the afterglow light curve and together account for the observed correlations. Especially if the two regions have strongly differing magnetizations, they can dominate the total emission in different spectral regimes (which might account for the lack of correlation between decay slopes in optical and X-rays reported by \citealt{Li2012}).

\begin{table}
\centering
\begin{tabular}{rrrrrr}
\hline
& & thick & thick & thick & thick \\
& & FS & RS & FS & RS \\
& & ISM & ISM & wind & wind \\
\hline
$\left<m\right>$ & LTX & 0.9 & 1.3 & 0.9 & 0.8 \\
 & LTO & 0.9 & 0.8 & 2.0 & 1.3 \\
$ \left< \rho \right>$ & LTX & $10^{-2}$ & $10^{-2}$ & $10^{-4}$ & $10^{-3}$ \\
& LTO & $10^{-2}$ & $10^{-2}$ & $10^{-3}$ & $10^{-2}$ \\
& $E-T$ & 0.5 & 0.5 & 0.5 & 0.5 \\
\hline
\end{tabular}
\caption{Same as table \ref{population_table}, now using $\epsilon_B = 10^{-6}$. }
\label{population_B_table}
\end{table}

If the degree of magnetization is altered, going from $\epsilon_B = 10^{-2}$ to a far lower $10^{-6}$ (see e.g. \citealt{Santana2013} for studies yielding significantly lower afterglow magnetizations than $\epsilon_B \sim 10^{-2}$), the results are as tabulated in table \ref{population_B_table}. The odds of chance correlations increase, but genuine correlations remain likely. On the whole, it becomes slightly harder to reproduce the LTX / LTO correlations from the literature, leading to sample based correlation results and literature LTX / LTO correlations that are sometimes consistent only within their $2 \sigma$ error bars. On the one hand, this indicates that, if a combination of FS and RS emission with different magnetizations is used to explain plateaus, the amount of freedom within parameter space is limited. On the other hand, the fact that going from $\epsilon_B = 10^{-2}$ to an extreme $\epsilon_B = 10^{-6}$ still leaves the thick shell model viable as an explanation for the LTX / LTO correlations at all, is an indication of the robustness of this result.

\section*{Acknowledgments}

I wish to thank Paul Duffell, Jochen Greiner, Alexander van der Horst, Andrew MacFadyen, Rafaella Margutti and Patricia Schady for helpful discussion and comments. This research was supported by the Alexander von Humboldt foundation.

\bibliographystyle{mn2e}
\bibliography{correlations}

\begin{thebibliography}{59}
\expandafter\ifx\csname natexlab\endcsname\relax\def\natexlab#1{#1}\fi

\bibitem[{{Achterberg} {et~al}\mbox{.}(2001){Achterberg}, {Gallant}, {Kirk}, \&
  {Guthmann}}]{Achterberg2001}
{Achterberg} A., {Gallant} Y.~A., {Kirk} J.~G., {Guthmann} A.~W., 2001, \mnras,
  328, 393

\bibitem[{{Berger} {et~al}\mbox{.}(2003){Berger}, {Kulkarni}, {Pooley},
  {Frail}, {McIntyre}, {Wark}, {Sari}, {Soderberg}, {Fox}, {Yost}, \&
  {Price}}]{Berger2003}
{Berger} E. {et~al.}, 2003, \nat, 426, 154

\bibitem[{{Cenko} {et~al}\mbox{.}(2011){Cenko}, {Frail}, {Harrison}, {Haislip},
  {Reichart}, {Butler}, {Cobb}, {Cucchiara}, {Berger}, {Bloom}, {Chandra},
  {Fox}, {Perley}, {Prochaska}, {Filippenko}, {Glazebrook}, {Ivarsen},
  {Kasliwal}, {Kulkarni}, {LaCluyze}, {Lopez}, {Morgan}, {Pettini}, \&
  {Rana}}]{Cenko2011}
{Cenko} S.~B. {et~al.}, 2011, \apj, 732, 29

\bibitem[{{Curran} {et~al}\mbox{.}(2009){Curran}, {Starling}, {van der Horst},
  \& {Wijers}}]{Curran2009}
{Curran} P.~A., {Starling} R.~L.~C., {van der Horst} A.~J., {Wijers}
  R.~A.~M.~J., 2009, \mnras, 395, 580

\bibitem[{{Dado} \& {Dar}(2013)}]{DadoDar2013}
{Dado} S., {Dar} A., 2013, \apj, 775, 16

\bibitem[{{Dai} \& {Lu}(1998)}]{Dai1998}
{Dai} Z.~G., {Lu} T., 1998, \aap, 333, L87

\bibitem[{{Dainotti}, {Cardone} \& {Capozziello}(2008){Dainotti}, {Cardone}, \&
  {Capozziello}}]{Dainotti2008}
{Dainotti} M.~G., {Cardone} V.~F., {Capozziello} S., 2008, \mnras, 391, L79

\bibitem[{{Dainotti} {et~al}\mbox{.}(2011){Dainotti}, {Fabrizio Cardone},
  {Capozziello}, {Ostrowski}, \& {Willingale}}]{Dainotti2011systematics}
{Dainotti} M.~G., {Fabrizio Cardone} V., {Capozziello} S., {Ostrowski} M.,
  {Willingale} R., 2011, \apj, 730, 135

\bibitem[{{Dainotti}, {Ostrowski} \& {Willingale}(2011){Dainotti}, {Ostrowski},
  \& {Willingale}}]{Dainotti2011}
{Dainotti} M.~G., {Ostrowski} M., {Willingale} R., 2011, \mnras, 418, 2202

\bibitem[{{Dainotti} {et~al}\mbox{.}(2013){Dainotti}, {Petrosian}, {Singal}, \&
  {Ostrowski}}]{Dainotti2013}
{Dainotti} M.~G., {Petrosian} V., {Singal} J., {Ostrowski} M., 2013, \apj, 774,
  157

\bibitem[{{Dainotti} {et~al}\mbox{.}(2010){Dainotti}, {Willingale},
  {Capozziello}, {Fabrizio Cardone}, \& {Ostrowski}}]{Dainotti2010}
{Dainotti} M.~G., {Willingale} R., {Capozziello} S., {Fabrizio Cardone} V.,
  {Ostrowski} M., 2010, \apjl, 722, L215

\bibitem[{{Duncan} \& {Thompson}(1992)}]{DuncanThompson1992}
{Duncan} R.~C., {Thompson} C., 1992, \apjl, 392, L9

\bibitem[{{Eichler} \& {Granot}(2006)}]{EichlerGranot2006}
{Eichler} D., {Granot} J., 2006, \apjl, 641, L5

\bibitem[{{Evans} {et~al}\mbox{.}(2009){Evans}, {Beardmore}, {Page}, {Osborne},
  {O'Brien}, {Willingale}, {Starling}, {Burrows}, {Godet}, {Vetere}, {Racusin},
  {Goad}, {Wiersema}, {Angelini}, {Capalbi}, {Chincarini}, {Gehrels}, {Kennea},
  {Margutti}, {Morris}, {Mountford}, {Pagani}, {Perri}, {Romano}, \&
  {Tanvir}}]{Evans2009}
{Evans} P.~A. {et~al.}, 2009, \mnras, 397, 1177

\bibitem[{{Filgas} {et~al}\mbox{.}(2011){Filgas}, {Kr{\"u}hler}, {Greiner},
  {Rau}, {Palazzi}, {Klose}, {Schady}, {Rossi}, {Afonso}, {Antonelli},
  {Clemens}, {Covino}, {D'Avanzo}, {K{\"u}pc{\"u} Yolda{\c s}}, {Nardini},
  {Nicuesa Guelbenzu}, {Olivares}, {Updike}, \& {Yolda{\c s}}}]{Filgas2011}
{Filgas} R. {et~al.}, 2011, \aap, 526, A113

\bibitem[{{Frail} {et~al}\mbox{.}(2001){Frail}, {Kulkarni}, {Sari},
  {Djorgovski}, {Bloom}, {Galama}, {Reichart}, {Berger}, {Harrison}, {Price},
  {Yost}, {Diercks}, {Goodrich}, \& {Chaffee}}]{Frail2001}
{Frail} D.~A. {et~al.}, 2001, \apjl, 562, L55

\bibitem[{{Gao} {et~al}\mbox{.}(2013){Gao}, {Lei}, {Zou}, {Wu}, \&
  {Zhang}}]{Gao2013}
{Gao} H., {Lei} W.-H., {Zou} Y.-C., {Wu} X.-F., {Zhang} B., 2013, \nar, 57, 141

\bibitem[{{Gehrels} {et~al}\mbox{.}(2004){Gehrels}, {Chincarini}, {Giommi},
  {Mason}, {Nousek}, {Wells}, {White}, {Barthelmy}, {Burrows}, {Cominsky},
  {Hurley}, {Marshall}, {M{\'e}sz{\'a}ros}, {Roming}, {Angelini}, {Barbier},
  {Belloni}, {Campana}, {Caraveo}, {Chester}, {Citterio}, {Cline}, {Cropper},
  {Cummings}, {Dean}, {Feigelson}, {Fenimore}, {Frail}, {Fruchter}, {Garmire},
  {Gendreau}, {Ghisellini}, {Greiner}, {Hill}, {Hunsberger}, {Krimm},
  {Kulkarni}, {Kumar}, {Lebrun}, {Lloyd-Ronning}, {Markwardt}, {Mattson},
  {Mushotzky}, {Norris}, {Osborne}, {Paczynski}, {Palmer}, {Park}, {Parsons},
  {Paul}, {Rees}, {Reynolds}, {Rhoads}, {Sasseen}, {Schaefer}, {Short},
  {Smale}, {Smith}, {Stella}, {Tagliaferri}, {Takahashi}, {Tashiro},
  {Townsley}, {Tueller}, {Turner}, {Vietri}, {Voges}, {Ward}, {Willingale},
  {Zerbi}, \& {Zhang}}]{Gehrels2004}
{Gehrels} N. {et~al.}, 2004, \apj, 611, 1005

\bibitem[{{Granot}, {K{\"o}nigl} \& {Piran}(2006){Granot}, {K{\"o}nigl}, \&
  {Piran}}]{Granot2006}
{Granot} J., {K{\"o}nigl} A., {Piran} T., 2006, \mnras, 370, 1946

\bibitem[{{Granot} \& {Kumar}(2006)}]{GranotKumar2006}
{Granot} J., {Kumar} P., 2006, \mnras, 366, L13

\bibitem[{{Granot} \& {Sari}(2002)}]{Granot2002}
{Granot} J., {Sari} R., 2002, \apj, 568, 820

\bibitem[{{Greiner} {et~al}\mbox{.}(2011){Greiner}, {Kr{\"u}hler}, {Klose},
  {Afonso}, {Clemens}, {Filgas}, {Hartmann}, {K{\"u}pc{\"u} Yolda{\c s}},
  {Nardini}, {Olivares E.}, {Rau}, {Rossi}, {Schady}, \&
  {Updike}}]{Greiner2011}
{Greiner} J. {et~al.}, 2011, \aap, 526, A30

\bibitem[{{Grupe} {et~al}\mbox{.}(2013){Grupe}, {Nousek}, {Veres}, {Zhang}, \&
  {Gehrels}}]{Grupe2013}
{Grupe} D., {Nousek} J.~A., {Veres} P., {Zhang} B.-B., {Gehrels} N., 2013,
  \apjs, 209, 20

\bibitem[{{Hascoet}, {Daigne} \& {Mochkovitch}(2014){Hascoet}, {Daigne}, \&
  {Mochkovitch}}]{Hascoet2014}
{Hascoet} R., {Daigne} F., {Mochkovitch} R., 2014, ArXiv: 1401.0751

\bibitem[{{Kirk} {et~al}\mbox{.}(2000){Kirk}, {Guthmann}, {Gallant}, \&
  {Achterberg}}]{Kirk2000}
{Kirk} J.~G., {Guthmann} A.~W., {Gallant} Y.~A., {Achterberg} A., 2000, \apj,
  542, 235

\bibitem[{{Kobayashi}, {Piran} \& {Sari}(1999){Kobayashi}, {Piran}, \&
  {Sari}}]{Kobayashi1999}
{Kobayashi} S., {Piran} T., {Sari} R., 1999, \apj, 513, 669

\bibitem[{{Kobayashi} \& {Sari}(2000)}]{KobayashiSari2000}
{Kobayashi} S., {Sari} R., 2000, \apj, 542, 819

\bibitem[{{Kocevski} \& {Butler}(2008)}]{Kocevski2008}
{Kocevski} D., {Butler} N., 2008, \apj, 680, 531

\bibitem[{{Leventis}, {Wijers} \& {van der Horst}(2014){Leventis}, {Wijers}, \&
  {van der Horst}}]{Leventis2014}
{Leventis} K., {Wijers} R.~A.~M.~J., {van der Horst} A.~J., 2014, \mnras, 437,
  2448

\bibitem[{{Li} {et~al}\mbox{.}(2012){Li}, {Liang}, {Tang}, {Chen}, {Xi},
  {L{\"u}}, {Gao}, {Zhang}, {Zhang}, {Yi}, {Lu}, {L{\"u}}, \& {Wei}}]{Li2012}
{Li} L. {et~al.}, 2012, \apj, 758, 27

\bibitem[{{Liang}, {Zhang} \& {Zhang}(2007){Liang}, {Zhang}, \&
  {Zhang}}]{Liang2007}
{Liang} E.-W., {Zhang} B.-B., {Zhang} B., 2007, \apj, 670, 565

\bibitem[{{MacFadyen} \& {Woosley}(1999)}]{MacFadyen1999}
{MacFadyen} A.~I., {Woosley} S.~E., 1999, \apj, 524, 262

\bibitem[{{Margutti} {et~al}\mbox{.}(2013){Margutti}, {Zaninoni}, {Bernardini},
  {Chincarini}, {Pasotti}, {Guidorzi}, {Angelini}, {Burrows}, {Capalbi},
  {Evans}, {Gehrels}, {Kennea}, {Mangano}, {Moretti}, {Nousek}, {Osborne},
  {Page}, {Perri}, {Racusin}, {Romano}, {Sbarufatti}, {Stafford}, \&
  {Stamatikos}}]{Margutti2013}
{Margutti} R. {et~al.}, 2013, \mnras, 428, 729

\bibitem[{{Morsony}, {Lazzati} \& {Begelman}(2007){Morsony}, {Lazzati}, \&
  {Begelman}}]{Morsony2007}
{Morsony} B.~J., {Lazzati} D., {Begelman} M.~C., 2007, \apj, 665, 569

\bibitem[{{Nousek} {et~al}\mbox{.}(2006){Nousek}, {Kouveliotou}, {Grupe},
  {Page}, {Granot}, {Ramirez-Ruiz}, {Patel}, {Burrows}, {Mangano}, {Barthelmy},
  {Beardmore}, {Campana}, {Capalbi}, {Chincarini}, {Cusumano}, {Falcone},
  {Gehrels}, {Giommi}, {Goad}, {Godet}, {Hurkett}, {Kennea}, {Moretti},
  {O'Brien}, {Osborne}, {Romano}, {Tagliaferri}, \& {Wells}}]{Nousek2006}
{Nousek} J.~A. {et~al.}, 2006, \apj, 642, 389

\bibitem[{{Panaitescu} \& {Kumar}(2000)}]{PanaitescuKumar2000}
{Panaitescu} A., {Kumar} P., 2000, \apj, 543, 66

\bibitem[{{Panaitescu} \& {Kumar}(2001)}]{PanaitescuKumar2001}
{Panaitescu} A., {Kumar} P., 2001, \apjl, 560, L49

\bibitem[{{Panaitescu} \& {Vestrand}(2008)}]{PanaitescuVestrand2008}
{Panaitescu} A., {Vestrand} W.~T., 2008, \mnras, 387, 497

\bibitem[{{Panaitescu} \& {Vestrand}(2011)}]{PanaitescuVestrand2011}
{Panaitescu} A., {Vestrand} W.~T., 2011, \mnras, 414, 3537

\bibitem[{{Peng}, {K{\"o}nigl} \& {Granot}(2005){Peng}, {K{\"o}nigl}, \&
  {Granot}}]{Peng2005}
{Peng} F., {K{\"o}nigl} A., {Granot} J., 2005, \apj, 626, 966

\bibitem[{{Racusin} {et~al}\mbox{.}(2009){Racusin}, {Liang}, {Burrows},
  {Falcone}, {Sakamoto}, {Zhang}, {Zhang}, {Evans}, \& {Osborne}}]{Racusin2009}
{Racusin} J.~L. {et~al.}, 2009, \apj, 698, 43

\bibitem[{{Racusin} {et~al}\mbox{.}(2011){Racusin}, {Oates}, {Schady},
  {Burrows}, {de Pasquale}, {Donato}, {Gehrels}, {Koch}, {McEnery}, {Piran},
  {Roming}, {Sakamoto}, {Swenson}, {Troja}, {Vasileiou}, {Virgili},
  {Wanderman}, \& {Zhang}}]{Racusin2011}
{Racusin} J.~L. {et~al.}, 2011, \apj, 738, 138

\bibitem[{{Ramirez-Ruiz}, {Celotti} \& {Rees}(2002){Ramirez-Ruiz}, {Celotti},
  \& {Rees}}]{RamirezRuiz2002}
{Ramirez-Ruiz} E., {Celotti} A., {Rees} M.~J., 2002, \mnras, 337, 1349

\bibitem[{{Ryan}, {Van Eerten} \& {MacFadyen}(2014){Ryan}, {Van Eerten}, \&
  {MacFadyen}}]{Ryan2014}
{Ryan} G., {Van Eerten} H., {MacFadyen} A., 2014, Manuscript in preparation

\bibitem[{{Santana}, {Barniol Duran} \& {Kumar}(2013){Santana}, {Barniol
  Duran}, \& {Kumar}}]{Santana2013}
{Santana} R., {Barniol Duran} R., {Kumar} P., 2013, ArXiv e-prints: 1309.3277

\bibitem[{{Sari} \& {Piran}(1995)}]{SariPiran1995}
{Sari} R., {Piran} T., 1995, \apjl, 455, L143

\bibitem[{{Sari}, {Piran} \& {Narayan}(1998){Sari}, {Piran}, \&
  {Narayan}}]{Sari1998}
{Sari} R., {Piran} T., {Narayan} R., 1998, \apjl, 497, L17+

\bibitem[{{Soderberg} {et~al}\mbox{.}(2006){Soderberg}, {Berger}, {Kasliwal},
  {Frail}, {Price}, {Schmidt}, {Kulkarni}, {Fox}, {Cenko}, {Gal-Yam}, {Nakar},
  \& {Roth}}]{Soderberg2006}
{Soderberg} A.~M. {et~al.}, 2006, \apj, 650, 261

\bibitem[{{Usov}(1992)}]{Usov1992}
{Usov} V.~V., 1992, \nat, 357, 472

\bibitem[{{van der Horst} {et~al}\mbox{.}(2014){van der Horst}, {Paragi}, {de
  Bruyn}, {Granot}, {Kouveliotou}, {Wiersema}, {Starling}, {Curran}, {Wijers},
  {Rowlinson}, {Anderson}, {Fender}, {Yang}, \& {Strom}}]{vanderHorst2014}
{van der Horst} A.~J. {et~al.}, 2014, MNRAS accepted. ArXiv e-prints: 1404.1945

\bibitem[{{Van Eerten}(2014)}]{vanEerten2014energyinjection}
{Van Eerten} H., 2014, Submitted. ArXiv: 1402.5162

\bibitem[{{Van Eerten} \& {Wijers}(2009)}]{vanEerten2009}
{Van Eerten} H.~J., {Wijers} R.~A.~M.~J., 2009, \mnras, 394, 2164

\bibitem[{{Wijers}, {Rees} \& {Meszaros}(1997){Wijers}, {Rees}, \&
  {Meszaros}}]{Wijers1997}
{Wijers} R.~A.~M.~J., {Rees} M.~J., {Meszaros} P., 1997, \mnras, 288, L51

\bibitem[{{Woosley}(1993)}]{Woosley1993}
{Woosley} S.~E., 1993, \apj, 405, 273

\bibitem[{{Yi}, {Wu} \& {Dai}(2013){Yi}, {Wu}, \& {Dai}}]{Yi2013}
{Yi} S.-X., {Wu} X.-F., {Dai} Z.-G., 2013, \apj, 776, 120

\bibitem[{{Zhang} {et~al}\mbox{.}(2006){Zhang}, {Fan}, {Dyks}, {Kobayashi},
  {M{\'e}sz{\'a}ros}, {Burrows}, {Nousek}, \& {Gehrels}}]{ZhangBing2006}
{Zhang} B., {Fan} Y.~Z., {Dyks} J., {Kobayashi} S., {M{\'e}sz{\'a}ros} P.,
  {Burrows} D.~N., {Nousek} J.~A., {Gehrels} N., 2006, \apj, 642, 354

\bibitem[{{Zhang} {et~al}\mbox{.}(2007){Zhang}, {Liang}, {Page}, {Grupe},
  {Zhang}, {Barthelmy}, {Burrows}, {Campana}, {Chincarini}, {Gehrels},
  {Kobayashi}, {M{\'e}sz{\'a}ros}, {Moretti}, {Nousek}, {O'Brien}, {Osborne},
  {Roming}, {Sakamoto}, {Schady}, \& {Willingale}}]{ZhangBing2007}
{Zhang} B. {et~al.}, 2007, \apj, 655, 989

\bibitem[{{Zhang} \& {M{\'e}sz{\'a}ros}(2001)}]{ZhangMeszaros2001}
{Zhang} B., {M{\'e}sz{\'a}ros} P., 2001, \apjl, 552, L35

\bibitem[{{Zhang}, {Woosley} \& {MacFadyen}(2003){Zhang}, {Woosley}, \&
  {MacFadyen}}]{ZhangWeiqun2003}
{Zhang} W., {Woosley} S.~E., {MacFadyen} A.~I., 2003, \apj, 586, 356

\end{thebibliography}

\bsp

\label{lastpage}

\end{document}